\documentclass[12pt]{article}

\usepackage{hyperref}
\usepackage{pdflscape}

\usepackage{graphics,psfrag}
\usepackage{amsthm,amssymb,epsfig,amsmath,euscript,array,cite}
\usepackage{color}
\usepackage{graphicx}
\usepackage{amscd}
\usepackage{slashed}
\usepackage{setspace}

\newcommand{\be}{\begin{equation}}
\newcommand{\ee}{\end{equation}}
\def\bea{\begin{eqnarray}}
\def\eea{\end{eqnarray}}

\def\nn{\nonumber}

\newcommand{\beq}{\begin{equation}}
\newcommand{\eeq}{\end{equation}}
\newcommand{\ben}{\begin{eqnarray}}
\newcommand{\een}{\end{eqnarray}}
\newcommand{\bes}{\begin{subequations}}
\newcommand{\ees}{\end{subequations}}
\newcommand{\blg}{\begin{align}}
\newcommand{\elg}{\end{align}}

\def\one{\mbox{1 \kern-.59em {\rm l}}}
\def\dphi1{{\dot\phi_1}}
\def\dphi2{{\dot\phi_2}}
\def\dphi3{{\dot\phi_3}}
\def\dphi{{\dot\phi}}

\def\={\, =\, }
\def\e{\epsilon}

\makeatletter \@addtoreset{equation}{section} \makeatother

\begin{document}

\begin{titlepage}

\hfill MCTP-15-19
\begin{center}

{\Large \bf Supergravity Solutions with $AdS_4$}\\

\vspace{.4cm}

{\Large \bf from Non-Abelian T-Dualities }

\vspace{.4cm}

{\bf \footnotesize  Leopoldo A. Pando Zayas$^{a,}$\footnote{lpandoz@umich.edu, $^2$vincent-rodgers@uiowa.edu, $^3$catherine.whiting@wits.ac.za}, Vincent G. J. Rodgers${^{b,2}}$, and Catherine A. Whiting$^{b,c,3}$}

\vspace{.2cm}

{\it ${}^a$ The Abdus Salam International Centre for Theoretical Physics}\\
{\it Strada Costiera 11, 34014 Trieste, Italy}\\

\vspace{.4cm}
{\it ${}^a$  Michigan Center for Theoretical Physics,  Department of Physics}\\
{\it University of Michigan, Ann Arbor, MI 48109, USA}\\

\vspace{.4cm}

{\it ${}^b$ Department of Physics and Astronomy}\\
{\it   The University of Iowa,   Iowa City, IA 52242, USA}\\

\vspace{.4cm}

{\it ${}^c$ National Institute for Theoretical Physics}\\
{\it School of Physics and Mandelstam Institute for Theoretical Physics}\\
{\it University of the Witwatersrand, Johannesburg}\\
{\it WITS 2050, South Africa}

\vspace{12pt}

\end{center}
\begin{abstract}
We present a large class of new backgrounds that are solutions of type II supergravity with a warped AdS${}_4$  factor, non-trivial axion-dilaton, B-field, and three- and five-form Ramond-Ramond fluxes. We obtain these solutions 
by applying non-Abelian T-dualities with respect to SU(2) or SU(2)/U(1) isometries to reductions to 10d IIA  of 11d sugra solutions of the form AdS${}_4 \times Y^7$, with  $Y^7 = S^7/\mathbb{Z}_k, S^7, M^{1,1,1}, Q^{1,1,1}$ and $N(1,1)$. The main class of reductions to IIA is along the Hopf fiber and leads to solutions of the form $AdS_4 \times K_6$, where $K_6 $ is K\"ahler  Einstein with $K_6=\mathbb{CP}^3, S^2\times \mathbb{CP}^2, S^2\times S^2 \times S^2$; the first member of this class is dual to the ABJM field theory in the 't Hooft limit. We also consider other less symmetric but susy preserving reductions along circles that are not the Hopf fiber.
In the case of $N(1,1)$ we find an additional breaking of isometries in the NAT-dual background.
To initiate the study of some properties of the field theory dual, we explicitly compute the central charge holographically.
\end{abstract}

\end{titlepage}

\setcounter{page}{1} \renewcommand{\thefootnote}{\arabic{footnote}}
\setcounter{footnote}{0}
\newpage

\onehalfspacing
\section{Introduction}

The AdS/CFT correspondence \cite{Maldacena:1997re} conjectures a complete equivalence between a field theory and a string theory. This correspondence provides an explicit recipe to connect observable quantities on both sides of the correspondence and it is particularly potent when the field theory is conformal \cite{Gubser:1998bc,Witten:1998qj,Aharony:1999ti}. A central conceptual and intuitive role is played by the geometerization of conformal invariance which on the gravity side is realized as the presence of an AdS factor whose isometries coincide with the conformal group of the theory living on the boundary.

More precisely, the conformal group of a field theory in $d$ dimensions is $SO(d,2)$ for $d>2$; this group is precisely the isometry group for AdS${}_{d+1}$. This relation  has prompted the search for solutions in type II supergravities and M-theory that contain AdS${}_{d+1}$ as a space-time  factor. Most of the effort has, naturally, been concentrated on searching for solutions with an AdS${}_5$ factor (see, for example, \cite{Gauntlett:2004zh}, \cite{Gauntlett:2005ww} and \cite{Colgain:2011hb}). There are, however, strong efforts in other dimensions. Recent efforts include AdS${}_7$  \cite{Apruzzi:2013yva}. In the context of theories connected to massive IIA, this search can be extended to AdS${}_6$ \cite{Apruzzi:2014qva,Kim:2015hya} and more recently to AdS${}_5$ \cite{Apruzzi:2015zna} and AdS${}_4$ \cite{Rota:2015aoa}. Another recent approach to the construction of gravity duals to conformal backgrounds has been presented in \cite{Gutowski:2014ova}, \cite{Beck:2014zda}, \cite{Beck:2015hpa}. There has also been recent progress in the direct search for solutions with AdS${}_{d+1}$ factors; interesting examples are the geometries interpolating between AdS${}_7$ and AdS${}_5$ presented in  \cite{Bah:2011vv,Bah:2012dg} and further studied and generalized in \cite{Bah:2013qya,Bah:2015fwa}.

Recently, there has been a revival of NATD  \cite{Ossa:1992vc,Fridling:1983ha,Fradkin:1984ai} including its systematic extension to the Ramond-Ramond sector \cite{Sfetsos:2010uq, Lozano:2011kb}.
This resurrected symmetry has already been used to generate solutions from various seed backgrounds in the context of the AdS/CFT correspondence \cite{Itsios:2013wd, Caceres:2014uoa, Lozano:2012au, Itsios:2012zv, Macpherson:2013lja, Gevorgyan:2013xka, Macpherson:2013zba, Lozano:2013oma, Gaillard:2013vsa, Elander:2013jqa, Zacarias:2014wta, Pradhan:2014zqa, Lozano:2014ata, Kelekci:2014ima, Macpherson:2014eza, Kooner:2014cqa, Araujo:2015npa, Bea:2015fja, Lozano:2015bra, Lozano:2015cra, Araujo:2015dba, Macpherson:2015tka, Dimov:2015rie}.
NATD is a powerful tool in the  search for solutions because it can be considered as a solution generating technique. Even more importantly, as shown in for example \cite{Macpherson:2014eza}, it leads to supersymmetric solutions with very few apparent isometries other than those of AdS.\footnote{In this paper we present an example where the NATD destroys even more isometries than the previously known examples.} This makes NATD a unique technique which is able to probe the space of solutions in directions that cannot be accessed by various general classification approaches which are based on explicit symmetries.

The main goal of this paper is to further use T-duality and NATD  to  construct supergravity backgrounds that contain an AdS${}_4$ factor in the metric. In  a very direct sense, this manuscript is an extension of \cite{Macpherson:2014eza} which was devoted to solutions with an AdS${}_5$ factor. There are various interesting applications for conformal field theories in $d=3$. For example, they have historically played an important role in our understanding of critical phenomena through the $\epsilon$ expansion of the Wilson-Fisher fixed point. More formally and recently,  3d conformal theories have helped clarify aspects of the M2 brane theory. We hope that the backgrounds presented in this manuscript will add some interesting new examples and motivate studies of the corresponding field theory duals which are arguably conformal or superconformal field theories in 3d. The quintessential example in this class is the ABJM field theory  \cite{Aharony:2008ug}; the NATD to the gravity background dual to the ABJM theory has been discussed explicitly in \cite{Lozano:2014ata}.

We summarize our strategy and main results in Figure \ref{Fig:generation}. Namely, we generate various families of 10d Type IIB solutions containing an AdS${}_4$ factor by first starting with a Freund-Rubin type solution in 11d supergravity of the form AdS${}_4\times Y^7$ and subsequently reduce the solution to IIA, which is followed by applying a non-Abelian T-duality to either an SU(2) or an SU(2)/U(1) isometry.  Generically, the final Type IIB backgrounds have all of the RR fluxes turned on, with the exception of the very symmetric case of $S^7$.  Finally, we compute the Page charges of these backgrounds and the central charge of the corresponding dual field theories.  We also investigate the change in the Page charges under large gauge transformations in $B_2$.

\begin{figure}[h!]
\begin{align*}
 \begin{CD}
 AdS_4 \times Y^7 @>Reduction>> \Downarrow(AdS_4\times Y^7) @>NATD >> \text{NATD}(\Downarrow(AdS_4\times Y^7)) \\
   \\
  F_4 @>Reduction>> (F_2, F_4) @>NATD  >> (F_1,F_3, F_5), H_3 \\
  \\
   Q_{D2}=N_{M2} @>Reduction>> Q_{D2}=N_{D2} @>NATD  >> \left(\begin{tabular}{c} $Q_{D5}=N_{D5}$\\ $\Delta Q_{D3}=nN_{D5}$\end{tabular}\right) \\ 
 \end{CD}
\end{align*}
\caption{\label{Fig:generation} A schematic description of some properties of the supergravity solutions discussed in this manuscript and their field theory duals.
}
\end{figure}

The paper is organized as follows. In Section \ref{Sec:General} we sketch the procedure that we apply to generate the various explicit solutions in the rest of the paper.  Sections \ref{Sec:S7/Zk}, \ref{Sec:S7}, \ref{Sec:M111}, \ref{Sec:Q111} and \ref{Sec:N11} correspond respectively to the application of our solution generating techniques to: $S^7/\mathbb{Z}_k, S^7, M^{1,1,1}, Q^{1,1,1} $ and $N(1,1)$. In Section \ref{Sec:CC}, we discuss our recipe for computing the central charge of the dual field theory holographically and present the results for one representative background from each section. We also propose a method for determining the ranges of the dual coordinates in backgrounds generated from NATD with coset space isometries, where $B_2$ is absent.  We discuss some arguments for the preservation of supersymmetry in some of the new backgrounds and conclude in Section \ref{Sec:conclusions}. We relegate to Appendix \ref{NATDReview} the rules for non-Abelian T-duality and Appendix \ref{ReviewEOM} the form of the 11d and 10d equations of motion as we used them to verify that the new backgrounds obtained in the manuscript are solutions.

\section{Freund-Rubin seeds for solution generating techniques using Non-Abelian T-duality}\label{Sec:General}

In this section we review a construction of solutions to 11-dimensional supergravity based on the Freund-Rubin Ansatz \cite{Freund:1980xh}. Essentially every Sasaki-Einstein 7d manifold provides a supersymmetric  solution  to 11d supergravity.  A fairly complete description of solutions of seven dimensional manifolds, providing Freund-Rubin solutions to 11d supegravity, was cataloged in Duff-Nilsson-Pope \cite{Duff:1986hr}.  The list includes further specification about those which are supersymmetric and states what fraction of the supersymmetry is preserved. An exhaustive list of Sasaki-Einstein seven-dimensional manifolds is presented in \cite{Friedrich:1990zg}. Here we focus on a particular but wide set of solutions, hoping to elucidate some of the most generic aspects for other $AdS_4$ solutions.

We start by considering Freund-Rubin type solutions to 11d supergravity,
\bea\label{Freund-Rubin}
ds^2&=&ds^2(AdS_4)+ds^2(Y^7), \nonumber \\
F&=&3 d\Omega_4,
\eea
where  $ds_7^2$ is an Einstein metric on $Y^7$.  In the case when  $Y^7$  is also Sasaki, the corresponding solution is supersymmetric.
In this manuscript we are particularly interested in solutions that are supersymmetric.
The metric $ds_4^2$ could also contain an asymptotically AdS${}_4$ black hole but we do not consider explicitly that case as it breaks supersymmetry.
If the metric on $Y^7$ admits a $U(1)$ isometry, which is always the case in the backgrounds we consider in this paper, it can be written as
\be
ds_7^2= a(y_n)(d\tau +{\cal A})^2 +ds^2(K_6),
\ee
where ${\cal A}$ is a connection on $K_6$ and the $y_n$ represent coordinates on $K_6$.
Then we can reduce the corresponding 11d solution to 10d Type IIA via the following rules:
\bea
\label{Reduction}
ds_{11}^2&=& e^{-\frac{2}{3}\Phi}ds_{10}^2 + e^{\frac{4}{3}\Phi}(d\tau +C_{(1)})^2, \nonumber \\[2mm]
C^{M}_{(3)}&=& C^{IIA}_{(3)}+B_{(2)}\wedge d\tau.
\eea
If one is concerned with the correct dimensional scalings while performing the reduction, it is important to note that the 11d backgrounds have $l_p$ as the natural length scale.  Therefore the dilaton must pick up this scale to compensate for the factors of the 11d radius, $\text{R}$, inherited from the $a(y_n)$.  This can be seen in the examples provided below.

Once we arrive at a Type IIA supergravity background using the prescription in Eq. (\ref{Reduction}), we can further apply a Non-Abelian T-duality (NATD) to the corresponding background, provided  that the appropriate symmetries are present. We will find that there are generically two cases where we can perform an NATD. In the first case there is a structure resembling an $S^3$ and in this case we apply the rules of NATD directly as in \cite{Sfetsos:2010uq,Itsios:2013wd}. In the second case we merely have one or more $S^2$ subspaces on which we may implement NATD by exploiting the coset structure of $S^2$ and following the prescriptions and results of \cite{Lozano:2011kb}.

Before we begin the presentation of the backgrounds, let us first define some notational conventions used, and clarify a few points about the NATD gauge fixing.   In all of the backgrounds presented in this paper, we choose a gauge fixing for NATD such that all three of the Lagrange multipliers are kept as coordinates in the dual background, i.e. $g=1$.  We will introduce them in terms of spherical polar coordinates, $(\theta_i,\phi_i,\psi)\to (\rho,\chi,\xi)$. We have also left arbitrary the gauge fixing constant that we label $m_2$, which arises from $v_i\to m_2 v_i$. We direct the reader to Appendix \ref{NATDReview} or \cite{Itsios:2013wd} for more on gauge fixing in NATD.  In Section \ref{Sec:CC} we will use the fact that $m_2$ may be absorbed into the definition of $\rho$.

Note that we will adopt the following conventions for $AdS_4$, $d\Omega_{AdS_4}$ and 2-spheres throughout this document:
\bea
ds^2(AdS_4)&=&\frac{r^2}{\text{R}^2}(-dx_0^2+dx_1^2+dx_2^2)+\frac{\text{R}^2}{r^2}dr^2,\nn \\
d\Omega_{AdS_4}&=&-\frac{r^2}{\text{R}^2}dr\wedge dx_0\wedge dx_1\wedge dx_2, \nn \\ ds^2(\Omega_2^{(i)})&=&d\theta_i^2+\sin^2\theta_i d\phi_i^2,\quad ds^2(\hat{\Omega}_2)=d\chi^2+\sin^2\chi d\xi^2, \nn \\
d\Omega_2^{(i)}&=&\sin\theta_i d\theta_i\wedge d\phi_i,\quad d\hat{\Omega}_2=\sin\chi d\chi\wedge d\xi.
\eea
\section{Background from $S^7/\mathbb{Z}_k$}
\label{Sec:S7/Zk}
We begin our presentation with the background $AdS_4\times S^7/\mathbb{Z}_k$, which is given by,
\be
ds^2=\frac{1}{4} ds^2(AdS_4)+ds^2(S^7/\mathbb{Z}_k).
\ee
The metric on $S^7/\mathbb{Z}_k$ is:
\bea
ds^2(S^7/\mathbb{Z}_k)&=&\lambda_2^2\Bigg[d\alpha^2+\cos^2\frac{\alpha}{2}ds^2(\Omega_2^{(1)})+\sin^2\frac{\alpha}{2}ds^2(\Omega_2^{(2)})\nn\\&~&+\sin^2\frac{\alpha}{2}\cos^2\frac{\alpha}{2}
(d\psi+\cos\theta_1\,d\phi_1-\cos\theta_2\,d\phi_2)^2\Bigg]+\frac{\lambda_3^2}{k^2}(d\tau+k^2{\cal A})^2,
\label{S7-metric}\nn \\
{\cal A}&=&\cos\alpha\,d\chi+\lambda_1\cos^2\frac{\alpha}{2}\cos\theta_1\,d\phi_1
+\lambda_1\sin^2\frac{\alpha}{2}\cos\theta_2\,d\phi_2.
\label{S7-form}
\eea
with $\lambda_1=2,\ \lambda_2=\frac{1}{2},\ \lambda_3=\frac{1}{4}$.
The angle $\tau$ parameterizes the $U(1)$ fiber over $\mathbb{CP}^3$, defined in the brackets of Eq.(\ref{S7-form}) above.  The ranges of the angles are
$0\leq\alpha,\theta_1,\theta_2\leq\pi$,
$0\leq\phi_1,\phi_2\leq2\pi$, $0\leq\psi\leq 4\pi$ and $0\leq\tau\leq\frac{2\pi}{k}$.  This geometry is supported by an $F_4=-\frac{3}{8\text{R}}\,d\Omega_{AdS_4}$.

After a reduction to Type IIA via $\tau$, we recover the familiar $AdS_4\times \mathbb{CP}^3$ background, which we label $ \Downarrow_{\tau}(AdS_4\times S^7/\mathbb{Z}_k)$ to match our labeling conventions.
In addition to the metric, the supergravity background has a dilaton, and  4-form and 2-form field strengths from the Ramond-Ramond (RR) sector,
\be
e^{2\Phi}=\frac{\text{R}^3\lambda_3^3}{l_p^3k^3}\,,
\qquad
F_4=-\frac{3}{8\text{R}}\,d\Omega_{AdS_4}\,,
\qquad
F_2=l_pk\,d{\cal A}\,.
\label{field-strengths}
\ee
Here $d\Omega_{AdS_4}$ is the volume form on $AdS_4$ and $F_2$ is
proportional to the K\"ahler form on $\mathbb{CP}^3$.
This IIA background is dual to a supersymmetric field theory with ${\cal N}=6$ susy in 3D; therefore the IIA background preserves the same number, 24, supercharges \cite{Aharony:2008ug}.

\subsection{NATD($\Downarrow_{\tau}(AdS_4\times S^7/\mathbb{Z}_k$))}
In this section, we perform an NATD along the ($\theta_1,\phi_1,\psi$) directions defining an SU(2) isometry.  This NATD was performed in \cite{Lozano:2014ata}, however, the NATD there acted along the $(\theta_2,\phi_2,\chi)$ SU(2) isometry.  It is true that the two NATD backgrounds are related by a simple map (in the notation of \cite{Lozano:2014ata} it is $\zeta\to \zeta+\frac{\pi}{2}$).  Our results differ slightly, particularly in the factors of k, due to the presence of the warp factor $e^{-\frac{2}{3}\Phi}$ left over from the reduction to IIA, absent from \cite{Lozano:2014ata}.   Our choice here leads to a slightly different structure of singularities in the resulting background.
\bea
\hat{ds}^2&=&\frac{\text{R}\lambda_3^2}{4kl_p}ds^2(AdS_4)+\frac{ \lambda _2^2 \lambda _3 \text{R}^3}{kl_p}  \left(\sin^2\frac{\alpha }{2}ds^2(\Omega_2^{(2)})+\mathit{d}\alpha ^2\right)\nn \\&~&+\frac{1}{4 \Delta  k^2 \alpha ' l_p^2}\bigg[ 4 k^2 \alpha '^2 l_p^2 m_2^4 \rho ^2 \mathit{d}\rho^2+4 \lambda _2^4 \lambda _3^2m_2^2 \text{R}^6 \cos ^4\frac{\alpha }{2}( \cos \chi\mathit{d}\rho  -\rho   \sin \chi\mathit{d}\chi )^2\nn \\&~&+\lambda_2^4 \lambda _3^2 m_2^2 \text{R}^6 \sin ^2\alpha  \cos^2\frac{\alpha }{2} \left(\sin ^2\chi  \left(\rho ^2\left(\mathit{d}\xi -\cos \theta _2\mathit{d}\phi _2 \right){}^2+\mathit{d}\rho ^2\right)\nn \right.\\&~&\left.+\rho ^2  \cos ^2\chi \mathit{d}\chi ^2 +\rho   \sin 2 \chi \mathit{d}\rho  \mathit{d}\chi  \right)\bigg], \nn \\
\hat{ B}_2&=&\frac{\lambda_2^2\lambda_3m_2\text{R}^3 }{4\Delta  k l_p}\bigg[\frac{ \sin ^2\alpha \cos \theta _2}{ k^2 \alpha '^2 l_p^2}\Big( \cos \chi  \left(k^2 m_2^2 \rho ^2 \alpha '^2 l_p^2+\lambda _2^4 \lambda _3^2 \text{R}^6 \cos ^4\frac{\alpha }{2}\right) \mathit{d}\rho\Big) \wedge\mathit{d}\phi _2 \nn \\&~&-\lambda _2^4 \lambda _3^2 \rho  \text{R}^6  \cos ^4\frac{\alpha}{2} \sin \chi  \mathit{d}\chi- m_2^2 \rho ^2 \sin \chi\left(\rho \Theta \mathit{d}\chi +  4\cos ^4\frac{\alpha }{2}  \cos \chi\mathit{d}\rho\right)\wedge  \mathit{d}\xi\bigg],\nn \\
 e^{-2\hat{\Phi}}&=&\left(\frac{kl_p}{\text{R}\lambda_3}\right)^3\Delta,
 \eea
where we have additionally defined,
\bea
\Delta=\frac{\lambda _2^2 \lambda _3 \text{R}^3 }{4 k^3 l_p^3 \alpha '^3}\left(k^2 l_p^2 m_2^2 \rho ^2 \alpha'^2\Theta+\lambda _2^4 \lambda _3^2  \text{R}^6 \sin ^2\alpha  \cos ^4\frac{\alpha }{2}\right), \text{and} \nn \\ \Theta=( \sin ^2\alpha  \cos ^2\chi +4\cos ^2\frac{\alpha }{2} \sin ^2\chi ),\ \Gamma=\left(4 k^2 m_2^2 \rho ^2 \alpha '^2 l_p^2+\lambda _2^4\lambda _3^2 \text{R}^6 \sin ^2\alpha \cos ^2\frac{\alpha }{2}\right).\nn
   \eea
   The RR sector contains the following field strengths:
\bea
\hat{F}_1&=&-\frac{k m_2 l_p \left(\lambda _1 \cos ^2\frac{\alpha }{2} (\rho  \sin\chi \mathit{d}\chi  -  \cos \chi \mathit{d}\rho)+\rho  \sin \alpha   \cos \chi\mathit{d}\alpha \right)}{\sqrt{\alpha '}}, \nn \\
\hat{F}_3&=&\frac{\lambda _2^6 \lambda _3^3 m_2^2 \rho ^2 \text{R}^9 \sin ^3\alpha  \cos ^4\frac{\alpha}{2}  \cos \chi }{4 \Delta  k^2 \alpha '^{5/2} l_p^2} \mathit{d}\alpha \wedge \big( \cos \theta _2\sin \chi  \mathit{d}\chi \wedge \mathit{d}\phi _2- d\hat{\Omega}_2\big) \nn \\&~&+\frac{\lambda _1 \lambda _2^2 \lambda _3 m_2^4 \rho ^3 \text{R}^3 \sin^2\alpha \cos ^2\frac{\alpha }{2}\cos\chi  }{4 \Delta \sqrt{\alpha '}}\mathit{d}\rho\wedge \big( \cos\theta _2 \sin \chi  \mathit{d}\chi \wedge \mathit{d}\phi _2- d\hat{\Omega}_2\big)\nn \\&~&+\frac{\lambda _2^2 \lambda _3 m_2^2 \rho  \text{R}^3 \sin \alpha  \cos ^2\frac{\alpha }{2}\sin ^2\chi  \Gamma}{4 \Delta k^2 \alpha '^{5/2} l_p^2} \mathit{d}\alpha \wedge \mathit{d}\rho \wedge\big(\cos \theta _2  \mathit{d}\phi _2-\mathit{d}\xi\big) \nn \\&~&+\big(k \lambda _1 m_2^2 \rho \sqrt{\alpha '} \sin ^2\frac{\alpha }{2} l_p\mathit{d}\rho +\frac{3 \lambda _2^6 \lambda _3 \text{R}^6  \sin^3\alpha   \mathit{d}\alpha}{4k \alpha'^{3/2} l_p}\big)\wedge d\Omega_2^{(2)},\nn \\
\hat{F}_5&=&\frac{\lambda_2^6 \lambda _3^2 m_2^3\text{R}^9\rho^2  \sin ^2\frac{\alpha}{2} \cos ^2\frac{\alpha }{2} \sin \alpha }{4 \Delta  k^2\alpha '^{3/2} l_p^2}\bigg[(\lambda _1  \lambda _3    \sin \alpha  \mathit{d}\rho -3 \lambda _2^2\rho    \Theta \mathit{d}\alpha  ) \wedge d\hat{\Omega}_2 \nn \\&~&-12 \lambda _2^2  \cos ^4\frac{\alpha }{2} \sin^2\chi  \cos \chi  \mathit{d}\alpha \wedge \mathit{d}\xi \wedge \mathit{d}\rho \bigg]\wedge d\Omega_2^{(2)}\nn \\&~& -d\Omega_{AdS_4}\wedge(\frac{\lambda _1 \lambda _2^2 \lambda _3^3 \text{R}^5 \sin \alpha  \cos ^2\frac{\alpha }{2} \mathit{d}\alpha }{32 k^2 \alpha '^{3/2} l_p^2}+\frac{3 m_2^2 \rho  \sqrt{\alpha '}\mathit{d}\rho }{8 \text{R}}).
\eea
The Einstein frame Ricci scalar for this background is (after setting $\text{R}=1,l_p=1,\alpha'=1,m_2=1, k=1$ for simplicity),
\bea
\mathcal{R}_E&=&\frac{1}{\left(4096 \rho ^2 \cos ^2\frac{\alpha  }{2} \cos 2 \chi +\left(2048 \rho ^2-1\right) \cos \alpha +2 \cos 2 \alpha +\cos 3 \alpha -6144 \rho ^2-2\right)^3}\bigg[\nn \\&~&4096\ 2^{3/4} \sec ^4\frac{\alpha }{2}\left(-\cos ^2\frac{\alpha }{2} \left(4096 \rho ^2 \cos ^2\frac{\alpha }{2} \cos 2 \chi +\left(2048 \rho ^2-1\right) \cos \alpha +2 \cos 2 \alpha\nn\right.\right. \\&~&\left.\left. +\cos 3 \alpha -6144 \rho ^2-2\right)\right)^{3/4} \left(65536 \rho ^4 (-\cos \alpha+\cos 2 \alpha -8)\nn \right.\\&~&-128 \rho ^2 (4 \cos \alpha +\cos 2 \alpha -21) \cos ^6\frac{\alpha }{2} \cos 2 \chi \nn \\&~&-32 \rho ^2 \cos ^4\frac{\alpha}{2} \left(-209 \cos \alpha +6 \cos 2 \alpha +\cos 3 \alpha -16384 \rho ^2 \cos 2\chi +266\right)\nn \\&~&\left.+\sin ^2\frac{\alpha }{2} (16 \cos \alpha +\cos 2 \alpha -39) \cos ^8\frac{\alpha }{2}\right)\bigg],
   \eea
which has singularities when $\alpha=\pi$, or when $\alpha=0$ simultaneous with $\rho=0$, or $\alpha=0$ simultaneous with $\chi=0$.  This can also be recognized, as in many examples of NATD, in the vanishing of $\Delta$, which is given by the $\det M$ (see Eq. (\ref{Mdef})).   These singularities are generated by the NATD, since the $AdS_4\times S^7/\mathbb{Z}_k$ background and its reduction to Type IIA are smooth.  NATD generates singularities whenever the duality is performed along a collapsing cycle, as in Abelian T-duality.

In \cite{Lozano:2014ata} the Killing spinor of $\mathbb{CP}^3$ was computed and shown to be dependent only on $\alpha$ ($\zeta$ in their coordinates) and two constants.  More precisely,
\cite{Lozano:2014ata} showed  that there are only 2 linearly independent Killing spinors on $\mathbb{CP}^3$ that are independent of the $SU(2)$  angles ($\theta_1,\phi_1,\psi$)
in the required frame.  Based on the demonstration given in \cite{Kelekci:2014ima}, which showed that supersymmetry is preserved when the Killing spinor of the original background is independent of the isometry direction, we can conclude that this NATD background preserves supersymmetry (at least $\mathcal{N}=2$ in 3d).
\section{Background from $AdS_4\times S^7$}
\label{Sec:S7}
In this section we consider a reduction to IIA along the $\psi_2$ direction of the $AdS_4\times S^7$ background, which we define as:
\bea
\label{S7}
ds_{11}^2&=&ds_{AdS_4}^2+ds_{S^7}^2, \nonumber \\
ds^2_7&=& d\mu^2 + \frac{1}{4}\sin^2\mu \omega_i^2 +\frac{1}{4}\lambda^2(\nu_i+\cos\mu \omega_i)^2, \nonumber \\
\nu_i&=& \sigma_i +\Sigma_i, \qquad \omega_i=\sigma_i -\Sigma_i\nn,\\
F_4&=&-\frac{3}{\text{R}}d\text{Vol}(AdS_4),
\eea
where $\sigma_i$ and $\Sigma_i$ are left-invariant $SU(2)$ Maurer Cartan 1-forms given by,
\bea
\sigma_1&=&-\sin\psi_1d\theta_1+\cos\psi\sin\theta_1d\phi_1,\nn\\
\sigma_2&=& \cos\psi_1 d\theta_1+\sin\psi\sin\theta_1d\phi_1, \nn \\
\sigma_3&=&\cos\theta_1 d\phi_1+d\psi_1,
\eea
and similarly for the $\Sigma_i$, but with coordinates ($\theta_2,\phi_2,\psi_2$).
The range of $\mu$ is $0\leq\mu\leq\pi$, as given in \cite{Duff:1986hr}.
This background is supersymmetric for $\lambda=1$, corresponding to the round $S^7$, and $\lambda= \frac{1}{\sqrt{5}}$ corresponding to the squashed $S^7$.  We will focus on the round $S^7$.
Here we perform the reduction along the U(1) angle $\psi_2$ defined in the $\sigma$'s above.\footnote{Note that it is only U(1) for the round $S^7$.  For the squashed $S^7$ only $\phi_1$ and $\phi_2$ are U(1) angles.}
The 10D background NS and RR sectors take the form:
\bea
ds_{10}^2&=&\frac{\text{R}}{l_p}\cos\frac{\mu}{2} \Big(ds^2(AdS_4)+\text{R}^2(\sin^2\frac{\mu}{2}\Sigma_i^2+\cos^2\frac{\mu}{2}ds^2(\Omega_2^{(2)})+d\mu^2)\Big), \nonumber\\
B_2&=&0,\quad e^{2\Phi}=\left(\frac{\text{R}}{l_p}\right)^3\cos^3\frac{\mu}{2}, \nonumber \\
C_1&=&l_p \cos\theta_2 d\phi_2,\quad F_2=-l_pd\Omega_2^{(2)},\nn \\
F_4&=&\frac{3}{\text{R}}d\text{Vol}AdS_4.
\eea
This Type IIA background has Einstein frame Ricci scalar, $\mathcal{R}_E=-\frac{3l_p^{1/4}(1+15\cos\mu)}{64\text{R}^{9/4}\cos^{9/4}{\frac{\mu}{2}}}$.
The singularity at $\mu=\pi$, however, can be understood by the presence of D6 branes. Indeed, the presence of the $F_2$ flux suggests a dual $C_7$ potential along the directions of $AdS_4$ and $\Sigma_i$. More importantly, one can introduce near $\mu=\pi$ the following coordinate: $(\pi-\mu) d\mu^2/2 = dr^2/\sqrt{r}$. In this new coordinate, near the singular point, the metric takes the form
\be
ds^2=\sqrt{r}\left(ds^2(AdS_4)+\Sigma_i^2\right)+\frac{1}{\sqrt{r}}\left(dr^2 +r^2 ds^2(\Omega_2^{(2))}\right),
\ee
which coincides, precisely, with the metric near a D6 brane source. Under the subsequent NATD transformations this singularity will persist, but we know its origin.


\subsection{NATD($\Downarrow_{\psi_2}(AdS_4\times S^7)$)}
We will now perform a NATD along the SU(2) isometry defined by the $\Sigma_i$.
The NATD metric, $B_2$, and dilaton have the following form,
\bea
\label{NATDofRoundS7}
\hat{ds}^2&=&\frac{\text{R}}{l_p}\cos\frac{\mu}{2}\Big(ds^2(AdS_4)+\text{R}^2(\cos^2\frac{\mu}{2}ds^2(\Omega_2^{(2)})+d\mu^2)\Big)\nn \\&~&+\frac{m_2^2\text{R}^6\rho^2\sin^2\mu\sin^2\frac{\mu}{2}}{4l_p^2\alpha'\Delta}ds^2(\tilde{\Omega})+\frac{2l_pm_2^2\alpha'^2}{\text{R}^3\sin\mu\sin\frac{\mu}{2}}d\rho^2, \nn \\
\hat{B}_2&=&\frac{\text{R}^3m_2^3\rho^3\sin\mu\sin\frac{\mu}{2}}{2l_p\Delta}\sin\chi d\xi\wedge d\chi, \quad
e^{-2\hat{\Phi}}=\frac{l_p^3\Delta}{\text{R}^3\cos^3\frac{\mu}{2}},\nn \\
\Delta&=&\frac{\text{R}^3\sin\mu\sin\frac{\mu}{2}}{2l_p^3\alpha'^3}(l_p^2m_2^2\alpha'^2\rho^2+\frac{1}{4}\text{R}^6\sin^2\mu\sin^2\frac{\mu}{2}).
\eea
The non-trivial dual RR Fluxes are,
\bea
\label{RR Fluxes Round S7}
\hat{F_3}&=&\big(3\frac{\text{R}^6}{8l_p\alpha'^{3/2}}\sin^3\mu d\mu+l_pm_2^2\sqrt{\alpha'}\rho d\rho\big)\wedge d\Omega_2^{(2)},\nn \\
\hat{F_5}&=&d\text{Vol}(AdS_4)\wedge\big[\frac{\text{R}^5\sin\mu\sin^2\frac{\mu}{2}}{2l_p^2\alpha'^{3/2}}d\mu+\frac{3m_2^2\sqrt{\alpha'}\rho}{\text{R}}d\rho\big] \\&~& -\frac{\text{R}^9m_2^2\rho^2\sin^3\mu\sin\frac{\mu}{2}}{16l_p^2\alpha'^{3/2}\Delta}\big(3\rho\sin\mu d\mu -2\sin^2\frac{\mu}{2}d\rho \big)\wedge d\Omega_2^{(2)}\wedge d\tilde{\Omega}_2\nn.
\eea
After this NATD, the Einstein frame Ricci scalar is,
\bea
\mathcal{R}_E&=&-\frac{l_p\alpha'^{3/4}(29+164\cos\mu+63\cos2\mu)}{32\text{R}^3\sin^{5/2}\mu(4l_p^2m_2^2\alpha'^2\rho^2+\text{R}^6\sin^2\frac{\mu}{2}\sin^2\mu)^{1/4}},\nn
\eea
where here we can again see that the singularities that appear correspond to the zeros of $\Delta$ ($\mu=0,\pi$), plus the singularity at $\mu=\pi$, inherited from the Type IIA background.
\section{Backgrounds from $AdS_4\times M^{1,1,1} $} \label{Sec:M111}
The space we are concerned with in this section is a $U(1)$ bundle over $\mathbb{CP}^2\times S^2$ with characteristic numbers $n_1$ and $n_2$ and metric given by,
\bea
\label{M111def}
ds^2&=&c_1^2ds^2(AdS_4)+\text{R}^2ds^2(M^{1,1,1}),\nn \\
ds^2(M^{1,1,1})&=&c_2^2\big(d\tau-n_1\sin^2\mu\sigma_3+n_2\cos\theta_1d\phi_1\big)^2+c_3^3(d\theta_1^2+\sin^2\theta_1d\phi_1^2)\nn \\&~&+c_4^2\big(d\mu^2+c_5^2\sin^2\mu(\sigma_1^2+\sigma_2^2+\cos^2\mu\sigma_3^2)\big),
\eea
where $c_1^2=\frac{3}{2\Lambda},\ c_2^2=\frac{3}{32\Lambda},\ c_3^2=\frac{3}{4\Lambda},\  c_4^2=\frac{9}{2\Lambda},\ c_5^2=\frac{1}{4}$. The supersymmetric case corresponds to $n_1=3, n_2=2$; in the notation used by Duff-Nilson-Pope \cite{Duff:1986hr} this space is naturally denoted by $M(-3,2)$. Here, however, we follow a slightly more modern notation widely used in the literature, $M^{1,1,1}$.
The corresponding 11D geometry has Ricci scalar $\mathcal{R}=-\frac{\Lambda}{\text{R}^2}$ and admits a 4-form Flux,
\be
C_3=c_1^3 \frac{r^3}{\text{R}^3}dx_0\wedge dx_1\wedge dx_2,\ \ \ \
F_4=\frac{3c_1^3}{\text{R}}d\text{Vol}(AdS_4),
\ee
and $0\leq \mu\leq \frac{\pi}{2}$.
In the IIA reduction along $\tau$, we obtain,
\bea
\label{Eq:CP2S2}
ds^2_{IIA}&=&\frac{c_2\text{R}^3}{l_p}\bigg(c_1^2ds^2(AdS_4)+c_4^2 (d\mu^2+c_5^2\sin^2\mu(\sigma_1^2+\sigma_2^2+\cos^2\mu\sigma_3^2))\nn \\&~&+c_3^2(d\theta_1^2+\sin^2\theta_1 d\phi_1^2)\bigg),\nn \\
B_2&=&0,\quad e^{2\Phi}=\frac{c_2^3\text{R}^3}{l_p^3}.
\eea
 This implies that, $C_3^{IIA}=C_3$, and $C_{(1)}=l_p(-n_1\sin^2\mu\sigma_3+n_2\cos\theta_1d\phi_1)$.
This reduced metric has Einstein frame Ricci Scalar, $\mathcal{R}_E=0$. Note that the metric on the six dimensional subspace, which we denote $\Downarrow_{\tau}( M^{1,1,1})$, is simply the product of $S^2\times \mathbb{CP}^2$. The solution is supported by the K\"ahler forms which appear as $dC_{(1)}$.  The field theory of this background was examined in \cite{Martelli:2008si}  where it was shown that this $\tau$ reduction of $AdS_4\times M^{1,1,1}$ is $\mathcal{N}=2$ supersymmetric.

\subsection{NATD($\Downarrow_{\tau}(AdS_4\times(M^{1,1,1}))$)}
\label{M111:SU(2)}
We are now poised to perform a Non-Abelian T-duality along the SU(2) isometry made explicit by the $SU(2)$ invariant one-forms $\sigma$.
\bea
\hat{ds}^2&=&\frac{c_2\text{R}^3}{l_p}(c_1^2ds^2(AdS_4)+c_3^2ds^2\Omega_2^{(1)}+c_4^2d\mu^2)\nn \\&~&+\frac{m_2^2}{\alpha ' l_p^2 \Delta } \left(c_2^2 c_4^4 c_5^4 \text{R}^6 \sin ^4\mu \left(\rho ^2 \mathit{d}\xi ^2 \cos ^2\mu  \sin^2\chi +\rho ^2 \mathit{d}\chi^2 \left(1-\sin ^2\mu \cos ^2\chi \right)\nn\right.\right. \\&~&\left.\left.+\mathit{d}\rho ^2 \left(1-\sin^2\mu  \sin ^2\chi \right)-2 \rho  \mathit{d}\rho \mathit{d}\chi  \sin ^2\mu  \sin \chi  \cos \chi\right)+m_2^2 \rho ^2 \alpha '^2\mathit{d}\rho ^2 l_p^2\right),\nn \\
\hat{B}&=&\frac{c_2 c_4^2 c_5^2 m_2^3 \rho ^2 \text{R}^3 \sin ^2\mu \sin \chi }{4 l_p \Delta}  \mathit{d}\xi \wedge  \left(\rho  \left(-2 \sin ^2\mu  \cos 2 \chi+\cos 2 \mu +3\right)\mathit{d}\chi\ \nn \right.\\&~&\left.-4 \sin ^2\mu  \sin \chi  \cos \chi  \mathit{d}\rho \right),\nn \\
e^{-2\hat{\Phi}}&=&\frac{l_p^3\Delta}{c_2^3\text{R}^3},  \\ \Delta&=&\frac{c_2 c_4^2 c_5^2 \text{R}^3 \sin ^2\mu }{\alpha '^3 l_p^3} \Big(\cos^2\mu  \left(c_2^2 c_4^4 c_5^4 \text{R}^6 \sin ^4\mu+m_2^2 \rho ^2 \alpha '^2 l_p^2 \cos^2\chi \right)\nn \\&~&+m_2^2 \rho ^2 \alpha '^2l_p^2 \sin ^2\chi \Big).\nn
\eea
The RR sector contains,
\bea
\hat{F}_1&=&\frac{m_2 n_1 l_p \sin ^2\mu (\rho  \mathit{d}\chi  \sin \chi -\cos \chi  (2 \rho  \mathit{d}\mu  \cot \mu +\mathit{d}\rho ))}{\sqrt{\alpha '}},\nn \\
   \hat{F}_3&=&\frac{1}{\alpha '^{3/2} l_p} \left(-\frac{3 c_2 c_3^2c_4^4 c_5^3 \text{R}^6 \sin ^3\mu  \cos \mu   }{c_1}\mathit{d}\mu-m_2^2 n_2 \rho \alpha '^2 l_p^2 \mathit{d}\rho \right)\wedge d\Omega_2^{(1)}\nn \\&~&+\frac{c_2 c_4^2 c_5^2 m_2^2 n_1 \rho ^2 \text{R}^3 \sin^4\mu  \cos ^2\mu  \sin \chi  \cos \chi}{\alpha '^{5/2}l_p^2 \Delta}  \left(2c_2^2 c_4^4 c_5^4 \text{R}^6 \sin ^3\mu  \cos \mu  \mathit{d}\mu \nn \right. \\&~&\left.-m_2^2 \rho  \alpha '^2 l_p^2 \mathit{d}\rho\right) \wedge \mathit{d}\xi \wedge\mathit{d}\chi \nn \\&~&+\frac{2 m_2^2 n_1 \rho  \sin \mu  \cos \mu  \sin ^2\chi \left(\alpha '^3 l_p^3\Delta-c_2^3 c_4^6 c_5^6 \text{R}^9 \sin^8\mu  \cos ^2\mu  \cos ^2\chi\right)}{\alpha '^{5/2} l_p^2 \left(\cos^2\mu  \cos ^2\chi +\sin ^2\chi \right)\Delta}d\mu\wedge d\xi\wedge d\rho, \nn \\
   \hat{F}_5&=& \frac{c_1^3}{c_3^2 \text{R} \alpha'^{3/2} l_p^2}d\text{Vol}(AdS_4)\wedge
 \left(3 c_3^2 m_2^2 \rho  \alpha'^2 l_p^2 \mathit{d}\rho +c_1 c_2^3c_4^4 c_5^3 n_2 \text{R}^6 \sin ^3\mu  \cos \mu \mathit{d}\mu \right) \\&~&
 +\frac{c_2^2c_4^6c_5^5m_2^3\rho^2R^9\sin^5\mu \cos\mu \sin\chi}{\alpha'^{3/2}l_p^2\Delta}\bigg(\frac{3c_3^3  \left(\sin^2\mu  \sin ^2\chi +\cos ^2\mu \right)}{c_1} d\mu \nn \\&~& - c_2c_5n_2\sin\mu\cos\mu d\rho  \bigg) d\xi\wedge d\chi\wedge d\Omega_2^{(1)}\nn \\&~& + \frac{3 c_2^4 c_3^2 c_4^{10} c_5^9 m_2^3 \rho ^2 \text{R}^{15} \sin^{11}\mu  \cos \mu  \cos \chi}{64 c_1 \alpha'^{15/2} l_p^8 \Delta^3}  \bigg(8c_2^2 c_4^4 c_5^4 \text{R}^6 \sin ^4\mu  \cos ^2\mu  \sin \chi \nn \\&~&+m_2^2 \rho ^2\alpha '^2 l_p^2 \Big((\cos 2 \mu +7) \sin \chi -2 \sin ^2\mu \sin 3 \chi \Big)\bigg)^2\mathit{d}\xi \wedge\mathit{d}\mu \wedge  \mathit{d}\rho \wedge d\Omega_2^{(1)}.\nn
  \eea
  This background has singularities at $\mu=0$ and $\mu=\frac{\pi}{2}$, simultaneous with $\rho=0$ or $\chi=0$, all generated by the NATD.
\subsection{NATD of SU(2)/U(1) isometry in $ \Downarrow_{\tau}(AdS_4\times(M^{1,1,1}))$}
\label{M111:S2}
After the reduction to IIA of $AdS_4\times M^{1,1,1}$  we have a six-dimensional Einstein-K\"ahler manifold $\Downarrow_{\tau}(M^{1,1,1})=\mathbb{CP}^2\times S^2$. In the previous section we performed an NATD along the  $SU(2)$ isometry present in $\mathbb{CP}^2$ (see its explicit form in Eq. (\ref{Eq:CP2S2})). Since $S^2$ is a coset manifold and there is a prescription for applying NATD on coset manifolds \cite{Lozano:2011kb}, we now proceed to apply this coset NATD along the $S^2$ factor.   We review this method in Appendix \ref{Coset Review}.  The dual NS fields are given by
 \bea
 \hat{ds}^2&=&\frac{c_2\text{R}^3}{l_p}\bigg(c_1^2ds^2(AdS_4)+c_4^2 (d\mu^2+c_5^2\sin^2\mu(\sigma_1^2+\sigma_2^2+\sigma_3^2))\bigg)\nn \\&~&+\frac{m_2^2l_p\alpha'^2}{c_2c_3^2\text{R}^3}(\frac{z}{\rho}dz+ d\rho)^2+\frac{c_2c_3^2\text{R}^3}{l_p\rho^2}dz^2,\nn \\
 \hat{B}&=&0,\quad  e^{-\hat{\Phi}}=\frac{m_2c_3l_p\rho}{c_2\sqrt{\alpha'}},
 \eea
 with the RR sector given by,
 \bea
\hat{F}_1&=&-\frac{m_2n_2l_p}{\sqrt{\alpha'}}dz,\nn \\
 \hat{F}_3&=&m_2^2n_1l_p\sqrt{\alpha'}z dz\wedge \Big(\sin^2\mu d\Omega_2^{(2)}-2\sin\mu\cos\mu d\mu\wedge \sigma_3\Big) ,  \nn \\
 \hat{F}_5&=&\frac{3m_2}{c_1l_p\sqrt{\alpha'}\text{R}^3}\bigg(-c_1^4m_2l_p\text{R}^2\alpha'd\text{Vol}(AdS_4)\wedge(zdz+\rho d\rho)\nn \\&~&+c_2c_3^2c_4^4c_5^3\text{R}^9\cos\mu\sin^3\mu dz\wedge d\mu\wedge d\psi\wedge d\Omega_2^{(2)}\bigg).
 \eea
This background has a singularity at $\rho=0$, as can be seen in the behavior of the dilaton.  This is generated by the collapsing cycle in Eq. (\ref{Eq:CP2S2}) given by $\theta_1\to 0$.
\subsection{Alternate Reduction: $\Downarrow_{\phi_1}(AdS_4\times M^{1,1,1})$}
In this section we present an alternate, supersymmetric reduction to Type IIA of $AdS_4\times M^{1,1,1}$ presented originally in \cite{Petrini:2009ur}.  We then perform an NATD on this background, thus providing a new supersymmetric Type IIB solution with an $AdS_4$ factor.

In \cite{Petrini:2009ur} it was shown that reducing along the U(1) $\phi_1$ angle yields a supersymmetric Type IIA background of the following form\footnote{using the notation in Eq. (\ref{M111def}) above},
\bea
ds^2&=&\sqrt{A}c_1^2ds^2(AdS_4)+ds_6^2,\nn \\
B&=&0,\quad e^{2\Phi}=\frac{A^{3/2}\text{R}^3}{l_p^3}
\eea
where
\be
\label{M111phi}
ds^2_6=\frac{\text{R}^3}{l_p}\sqrt{A}\big[c_4^2d\mu^2+c_3^2d\theta_1^2+c_4^2c_5^2\sin^2\mu(\sigma_1^2+\sigma_2^2+\cos^2\mu\sigma_3^2)+c_2^2c_3^2\sin^2\theta_1(d\tau-n_1\sin^2\mu\sigma_3)^2\big],
\ee
and $A=c_2^2n_2^2\cos^2\theta_1+c_3^2\sin^2\theta_1$.

The RR sector contains the $F_4=-\frac{3c_1^3}{\text{R}}d\text{Vol}(AdS_4)$ flux proportional to the differential volume on $AdS_4$, while the reduction produces an additional $C_1$ and $F_2$:
\bea
C_1&=&c_2^2n_2l_p\frac{\cos\theta_1}{A}(d\tau-n_1\sin^2\mu\sigma_3), \nn \\ F_2&=&\frac{c_2^2 n_2 l_p }{A^2}\bigg[n_1 \sin \mu  \left(A \cos \theta _1 (\sin \mu d\Omega_2^{(2)}-2 \cos \mu \mathit{d}\mu \wedge (\cos\theta_2 \mathit{d}\phi_2+ \mathit{d}\psi ))\nn \right.\\&~&\left.+W \sin \mu( d\Omega_2^{(2)}- \mathit{d}\psi \wedge\mathit{d}\theta _1)\right)+W \mathit{d}\tau \wedge \mathit{d}\theta_1\bigg].
\eea
\subsubsection{NATD($\Downarrow_{\phi_1}(AdS_4\times M^{1,1,1})$)}
Now we move on to present the Non-Abelian T-dual with respect to an SU(2) isometry given in the $\sigma$'s of Eq. (\ref{M111phi}) above.
\bea
\hat{ds}^2&=&\sqrt{A}c_1^2ds^2(AdS_4)+\frac{\text{R}^3}{l_p}\sqrt{A}\big[c_4^2d\mu^2+c_3^2d\theta_1^2+\frac{c_2^2c_3^2c_4^4c_5^2\sin^2\theta_1}{Q}d\tau^2\big]+\hat{ds}^2_3,\nn \\
\hat{ds}^2_3&=&\frac{m_2^2 \sin ^4\mu}{\alpha ' l_p^2}\left(\frac{c_4^2 c_5^2 \text{R}^3 \alpha '^3  \sqrt{A} l_p^3 \csc ^6\mu  \sec ^2\mu  ( \cos \chi\mathit{d}\rho  -\rho   \sin \chi \mathit{d}\chi )^2}{V}\nn \right.\\&~&+\frac{\left(c_4^2 c_5^2 Q \text{R}^6 \cos \mu  ( \sin \chi\mathit{d}\rho  +\rho   \cos \chi \mathit{d}\chi )+m_2^2 \rho ^2   \alpha '^2  l_p^2 \csc ^4\mu \sec \mu  \sin \chi \mathit{d}\rho\right){}^2}{\Delta  V}\nn \\&~&\left.+\frac{c_4^2 c_5^2 \rho ^2 \text{R}^6 \sec ^2\mu  \sin ^2\chi   \left(Q \cos ^2\mu \mathit{d}\xi  -c_2^2 c_3^2 n_1  \sin ^2\theta_1\mathit{d}\tau \right){}^2}{\Delta  Q}\right),\nn \\
          B&=&\frac{\sqrt{A} c_2^2 c_3^2 c_4^4 c_5^4 m_2 n_1 \rho  \text{R}^9 \sin ^2\theta _1 \sin ^6\mu \sin \chi  \mathit{d}\tau \wedge \mathit{d}\chi }{\Delta \alpha '^2  l_p^3}\nn \\&~&+\frac{c_2^2 c_3^2 m_2 n_1 \text{R}^3 \sin ^2\theta _1\sin ^2\mu  \cos\chi  \left(A c_4^4 c_5^4 \text{R}^6 \sin ^4\mu+m_2^2 \rho ^2 \alpha '^2 l_p^2\right)\mathit{d}\rho \wedge \mathit{d}\tau }{\sqrt{A} \Delta \alpha '^2 l_p^3} \nn \\&~&+\frac{m_2^3 \rho ^3 \text{R}^3 \sin ^2\mu  \sin ^3\chi  \left(A c_4^2 c_5^2+Q \cos ^2\mu  \cot ^2\chi \right)\mathit{d}\xi \wedge\mathit{d}\chi }{\sqrt{A} \Delta  l_p}\nn \\&~&+\frac{m_2^3 \rho ^2 \text{R}^3 \sin ^2\mu  \sin ^2\chi  \cos \chi \left(Q \cos ^2\mu -A c_4^2 c_5^2\right) \mathit{d}\xi\wedge \mathit{d}\rho }{\sqrt{A} \Delta  l_p}, \nn \\
e^{-2\Phi}&=& \frac{l_p^3}{\text{R}^3A^{3/2}}\tilde{\Delta},
   \eea
       with $\Delta=\frac{\text{R}^3 \sin ^2\mu}{\alpha '^3 \sqrt{A} l_p^3} \left(c_4^2 c_5^2 m_2^2 \rho ^2\alpha '^2 A l_p^2 \sin ^2\chi +Q \cos ^2\mu  \left(c_4^4 c_5^4 \text{R}^6 A \sin ^4\mu +m_2^2 \rho ^2 \alpha '^2l_p^2 \cos ^2\chi \right)\right)$,
       $Q=c_4^2 c_5^2A+c_2^2 c_3^2 n_1^2 \sin ^2\theta _1 \tan^2\mu $, $V=c_4^2 c_5^2 Q \text{R}^6+m_2^2 \rho ^2\alpha '^2 l_p^2 \csc ^4\mu  \sec ^2\mu  \sin ^2\chi $, and $W=A\sin\theta_1+A'\cos\theta_1$.  This NATD generates singularities at $\mu=0$ and $\mu=\frac{\pi}{2}$, with $\rho=0$, or $\chi=0$.

       The RR sector fields take the following form,\\
       \bea
       F_1&=&\frac{c_2^2 m_2 n_1 n_2}{A\sqrt{\alpha'}}\bigg( \frac{1}{A}\rho   l_p \sin^2\mu  \cos \chi W \mathit{d}\theta _1\nn \\&~&- \cos\theta _1 l_p \sin ^2\mu  \cos \chi \mathit{d}\rho+\rho  \cos\theta _1 l_p \sin ^2\mu  \sin \chi \mathit{d}\chi -\rho   \cos \theta _1  l_p \sin \mu  \cos \mu  \cos \chi \mathit{d}\mu\bigg), \nn \\
       F_3&=&\Big( \frac{3c_2c_3^2c_4^4c_5^3\text{R}^6}{c_1l_p\alpha'^{3/2}}\cos\mu\sin^3\mu\sin\theta_1 d\mu+\frac{c_2^2c_4^2c_5^2m_2^2n_2l_p\sqrt{\alpha'}}{A Q}W\rho d\rho\Big)\wedge d\theta_1\wedge d\tau\nn \\&~& +\frac{c_2^2m_2^4n_1n_2 \text{R}^3}{\sqrt{\alpha'}A^{3/2}\Delta M}\cos\theta_1\sin^4\mu\cos\chi\sin\chi\big(\cos^2\mu Q d\xi-c_2^2c_3^2n_1\sin^2\theta_1 d\tau\big) \wedge d\rho\wedge d\chi \nn \\&~&+\frac{c_2^2 c_4^2 c_5^2 m_2^2 n_1 n_2 \rho  \text{R}^3 W \sin ^4\mu  \sin \chi }{A^{3/2}  \Delta  \alpha '^{5/2} l_p^2}\left(\sin\chi   \left(c_4^2 c_5^2 Q \text{R}^6 \sin ^4\mu  \cos ^2\mu +m_2^2 \rho ^2 \alpha'^2 l_p^2\right) \mathit{d}\rho\nn \right.\\&~&+\left.c_4^2 c_5^2 \rho  Q \text{R}^6 \sin ^4\mu  \cos ^2\mu  \cos \chi  \mathit{d}\chi\right)\wedge \mathit{d}\xi \wedge \mathit{d}\theta_1\nn \\&~& -\frac{c_2^4 c_3^2 c_4^2 c_5^2 m_2^2 n_1^2 n_2 \rho  \text{R}^3 W \sin ^2\theta _1\sin ^2\mu  \sin \chi}{A^{3/2}\Delta  Q \alpha'^{5/2} l_p^2} \left(m_2^2\rho ^2 \alpha '^2 l_p^2 \tan ^2\mu  \sin \chi  \mathit{d}\rho \nn \right. \\&~&\left.+c_4^2 c_5^2 Q \text{R}^6 \sin ^6\mu  \left(\rho  \cos \chi \mathit{d}\chi+\sin \chi \mathit{d}\rho\right)\right) \wedge \mathit{d}\tau \wedge \mathit{d}\theta _1 \nn \\&~&+\frac{c_2^2 c_4^2 c_5^2 m_2^2 n_1 n_2 \rho  \text{R}^3 \cos \theta _1\sin ^3\mu \cos \mu }{\sqrt{A} \Delta \alpha '^{5/2} l_p^2} \left(c_4^2 c_5^2 \rho  Q \text{R}^6 \sin ^4\mu \cos ^2\mu \sin 2 \chi \mathit{d}\chi\nn\right. \\&~&\left.+2 \sin ^2\chi \left(c_4^2 c_5^2 Q \text{R}^6 \sin ^4\mu  \cos ^2\mu +m_2^2 \rho ^2\alpha '^2 l_p^2\right)\mathit{d}\rho\right)\wedge \mathit{d}\mu \wedge \mathit{d}\xi \nn \\&~&-\frac{2 c_2^4 c_3^2 m_2^2 n_1^2 n_2 \rho  \text{R}^3 \sin ^2\theta _1 \cos\theta _1 \sin ^3\mu  \cos \mu  \cos \chi  }{A^{3/2} \Delta \alpha '^{5/2} l_p^2} \left(A c_4^4 c_5^4 \rho  \text{R}^6 \sin^4\mu  \sin \chi   \mathit{d}\chi\nn \right.\\&~&\left.+\cos \chi  \left(A c_4^4 c_5^4 \text{R}^6 \sin ^4\mu+m_2^2 \rho ^2 \alpha '^2 l_p^2\right) \mathit{d}\rho\right)\wedge \mathit{d}\mu\wedge  \mathit{d}\tau, \nn \\
       F_5&=&  \frac{c_1^3}{\text{R}\alpha'^{3/2}} d\text{Vol}(AdS_4)\wedge \bigg[-\frac{\text{R}^6}{l_p^2}c_1c_2^3c_3^2c_5n_1^2n_2\sin2\theta_2  \sin^4\mu d\theta_1+3m_2^2\alpha'^2\rho  d\rho\nn \\&~&+\frac{\text{R}^6}{l_p^2}\frac{c_1c_2c_4^4c_5^3n_2}{c_3^2}\cos\mu\sin^3\mu \csc\theta_1 W d\mu\bigg]\nn \\&~& + \frac{c_2c_4^4c_5^3m_2^3\text{R}^9\cos\mu\sin^5\mu\sin\chi}{c_1l_p^2\alpha'^{3/2}\sqrt{A}\Delta}\bigg(c_4^2c_5^2A\sin\theta_1\Big( 3c_3^2\sin\chi d\theta_1\wedge d\tau\wedge d\xi\wedge [d\mu\wedge\nn \\&~&(\cos\chi d\rho+\rho\sin\chi d\chi)+c_1c_2c_5n_2\cos\mu\sin\mu d\rho\wedge d\chi ]\Big)\nn \\&~&+c_3^2\sin\theta_1\Big(3\cos^2\mu\cos\chi Q d\theta_1\wedge d\mu\wedge \xi\wedge d\tau\wedge (\sin\chi d\rho+\rho\cos\chi d\chi)\nn \\&~&+c_1c_2^3c_5n_1^2n_2\sin2\theta_1\sin^2\mu d\mu\wedge d\xi\wedge d\rho\wedge d\tau\wedge d\chi\Big)\nn \\&~&+c_1c_2c_4^2c_5^3n_2\cos\theta_1\cos\mu\sin\mu A' d\theta_1\wedge d\xi\wedge d\rho\wedge d\tau\wedge d\chi  \bigg).
       \eea
\section{Backgrounds from $AdS_4\times Q^{1,1,1}$}\label{Sec:Q111}
The $Q(n_1,n_2,n_3)$ spaces (Equation 9.2.12 in Duff-Nilsson-Pope) \cite{Duff:1986hr} are defined as:
\be
ds^2=c^2(d\tau + \sum\limits_{i=1}^3 \,\, n_i \cos\theta_i d\phi_i)^2+\sum\limits_{i=1}^3 \frac{1}{\Lambda_1}(d\theta_i^2+\sin^2\theta_i d\phi_i^2).
\ee
The $n_1=n_2=n_3=1$ is ${\cal N}=2$ supersymmetric. Aspects of the supersymmetry were first presented in \cite{DAuria:1983vy}.  In \cite{Franco:2009sp} the field theory dual to $AdS_4\times Q^{1,1,1}/\mathbb{Z}_k$ was studied and various gauge invariant chiral operators were matched geometrically.

The 11D metric on $AdS_4\times Q^{1,1,1}$ is defined such that
\be
ds^2=t_1^2ds^2(AdS_4)+\text{R}^2ds^2(Q^{1,1,1}),
\ee
where we define, $t_1^2=\frac{3}{2\Lambda},\ t_2^2=\frac{3}{4\Lambda},\ t_3^2=\frac{3}{8\Lambda}$, and
\bea
&ds^2_{Q^{1,1,1}}= t_3^2(d\tau-\mathcal{A})^2+t_2^2(ds^2(\Omega_2^{(1)})+ds^2(\Omega_2^{(2)})+ds^2(\Omega_2^{(3)})),
\eea
where $\mathcal{A}=(\alpha-\cos\theta_1)d\phi_1+(\beta-\cos\theta_2)d\phi_2+(\gamma-\cos\theta_3)d\phi_3$, and the $ds^2(\Omega_2^{(i)})$ are 2-spheres with coordinates $\theta_i, \phi_i$. The 11D Ricci scalar is $\mathcal{R}=-\frac{\Lambda}{L^2}$, and the $F_4$ flux is normalized according to $F_4=\frac{3t_1^3}{\text{R}}d\text{Vol}(AdS_4)$.
The reduction to IIA along the $\tau$ angle yields the metric of the form,
\bea
\label{Eq:Q111_6}
ds^2&=&\frac{\text{R}t_3}{l_p}\big[t_1^2ds^2(AdS_4)+t_2^2\text{R}^2(ds^2(\Omega_2^{(i)}))\big], \nn\\
e^{2\Phi}&=&\frac{\text{R}^3t_3^3}{l_p^3}, \quad C_1=l_p\mathcal{A},
\eea
along with the same $F_4$ carried through the reduction. The total Ricci scalar vanishes for this background, $\mathcal{R}=0$.

The NATD of the background in Eq. (\ref{Eq:Q111_6}) when dualizing along the $(SU(2)/U(1))^3$ isometry was found in \cite{Lozano:2011kb}.
Here we present the full results for dualizing along just one of the $S^2$'s, i.e. one $(SU(2)/U(1))$ isometry.  Furthermore, we pay attention to the factors of R, the 11D radius, $l_p$ and $\alpha'$ throughout the dualization.  We review general aspects of NATD with coset spaces in Appendix \ref{Coset Review} and refer the reader there for more details.
 We find that,\bea
d\hat{s}^2&=&\frac{t_1^2t_3\text{R}}{l_p}ds^2(AdS_4)+\bigg(\frac{l_p\alpha'^2m_2^2}{\text{R}^3t_2^2t_3\cos^2\chi}+\frac{\text{R}^3t_2^2t_3\tan\chi^2}{l_p\rho^2}\bigg)d\rho^2\nn \\&~&+\frac{\text{R}^3t_2^2t_3}{l_p}\bigg[ds^2(\Omega_2^{(1)})+ds^2(\Omega_2^{(2)})+d\chi^2+\frac{2\tan\chi}{\rho}d\rho d\chi\bigg],\nn  \\
\hat{B}&=&0,\quad e^{-\hat{\Phi}}=\frac{l_pm_2t_2\rho\cos\chi}{t_3\sqrt{\alpha'}}.
\eea
This background has Einstein frame Ricci scalar, $\mathcal{R}_E=\frac{\text{R}^6 t_1^2 t_2^4 t_3^2+\alpha'^2 l_p^2 m_2^2\rho ^2 \left(7 t_1^2-12t_2^2\right)}{2 \alpha'^{7/4} \text{R}^3 l_p^{3/2}m_2^{5/2} \rho ^{5/2} t_1^2 t_2^{5/2}\sqrt{t_3} \cos^{\frac{5}{2}}\chi}$.  The RR sector has all fields turned on,
\bea
\hat{F}_1&=&-\frac{l_p}{\sqrt{\alpha'}}m_2(\sin\chi d\rho+\rho \cos\chi d\chi),\nn \\
\hat{F}_3&=&l_p\sqrt{\alpha'}m_2^2\rho d\rho\wedge( d\Omega_2^{(1)} +d\Omega_2^{(2)}), \nn \\
\hat{F}_5&=&-\frac{3m_2t_1^3\sqrt{\alpha'}\rho}{\text{R}}d\text{Vol}(AdS_4)\wedge d\rho\nn \\&~&-\frac{3m_2t_2^6t_3\text{R}^6}{t_1l_p\sqrt{\alpha'}} (\sin\chi d\rho +\rho \cos\chi d\chi)\wedge d\Omega_2^{(1)}\wedge d\Omega_2^{(2)}.\nn
\eea
This background has singularities at $\rho=0$ and $\chi=0$, both generated by the NATD.
\subsection{Alternate Reduction: $\Downarrow_{\phi_i}(AdS_4\times Q^{1,1,1})$}
In this section, we perform a different reduction to Type IIA on the $AdS_4\times Q^{1,1,1}$ background, in such a way that two SU(2) isometries are manifest.  In order to see these isometries, we must set the parameters appearing in the K{\"a}hler form, $\alpha,\beta,\gamma$, to zero.  We must also make use of the relations between the $t_i$'s such that $t_1=2t_3$, and $t_2=\sqrt{2}t_3$. (NB: the equations of motion are satisfied for any value of $t_3$.)

Here we will present explicitly the reduction along $\phi_3$, however from the symmetry of $Q^{1,1,1}$ it is evident that reducing along any of the $\phi_i$ will yield backgrounds of the same general form.
\bea
ds^2&=&e^{\frac{2}{3}\Phi}\big[4t_3 ds^2(AdS_4)+\text{R}^22t_3^2\big(d\Omega_2^{(1)}+d\Omega_2^{(2)}+d\theta_3^2\nn \\&~&+\frac{ \sin^2\theta_3}{A}(\cos\theta_1d\phi_1+\cos\theta_2d\phi_2+d\psi)^2\big)  \big],\nn \\
B_2&=&0,\quad e^{2\Phi}=\frac{t_3^3\text{R}^3A^{3/2}}{2\sqrt{2}l_p^3},
\eea
where $A=3-\cos2\theta_3$.  We can rewrite the metric above in terms of the $\sigma_i$,
\bea
ds^2&=&e^{\frac{2}{3}\Phi}\big[4t_3 ds^2(AdS_4)+\text{R}^22t_3^2\big(d\Omega_2^{(1)}+\sigma_1^2+\sigma_2^2+d\theta_3^2\nn \\&~&+\frac{2\sin^2\theta_3}{A}(\cos\theta_1d\phi_1+\sigma_3)^2\big)  \big].
\eea
The RR sector contains,
\bea
C_1&=& \frac{l_p 2\cos\theta_3}{A}(\cos\theta_1 d\phi_1+\cos\theta_2d\phi_2+d\psi),\ F_2=dC_1\nn \\
F_4&=&-\frac{24t_3^3}{\text{R}}d\Omega_{AdS_4}.
\eea
\subsubsection{NATD($\Downarrow_{\phi_i}(AdS_4\times Q^{1,1,1})$)}
Now we will present the results for an NATD on the Type IIA background found in the previous section.  The NATD background we will present will be along the ($\theta_2,\phi_2,\psi$) SU(2) isometry, however the results from the ($\theta_1,\phi_2,\psi$) SU(2) isometry can easily be read off by replacing ($\theta_1\to \theta_2,\phi_1\to \phi_2$).  It is given by,
\bea
\hat{ds}^2&=&e^{\frac{2}{3}\Phi}\big[4t_3 ds^2(AdS_4)+\text{R}^22t_3^2\big(d\Omega_2^{(1)}+d\theta_3^2\big)\big]+\hat{ds}^2_3,\nn \\
\hat{ds}^2_3&=&\frac{m_2^2}{l_p^2\alpha'}\bigg[\frac{\left( \cos \chi  \left(2 A \text{R}^6 t_3^6+m_2^2 \rho ^2 \alpha '^2 l_p^2\right)\mathit{d}\rho -2 A \rho \text{R}^6 t_3^6  \sin \chi \mathit{d}\chi  \right){}^2}{\Delta \Xi}\nn \\&~&+\frac{\sqrt{2} \sqrt{A}\text{R}^3 t_3^3 \alpha '^3 l_p^3 ( \sin\chi \mathit{d}\rho +\rho   \cos \chi \mathit{d}\chi )^2}{\Xi}\nn \\&~&
+\frac{4 \rho ^2 \text{R}^6 t_3^6 \sin^2\theta _3 \sin ^2\chi \left(\cos \theta _1\mathit{d}\phi_1 +\mathit{d}\xi\right){}^2}{\Delta }  \bigg],\nn \\
\hat{B}_2&=&\frac{\sqrt{2}t_3^3m_2 \text{R}^3}{l_p\Delta\sqrt{A}}\bigg[m_2^2 \rho ^2 \Big(\rho \left(A \sin ^2\chi +2 \sin^2\theta _3 \cos ^2\chi\right)d\hat{\Omega}_2+2   \sin ^2\chi \cos \chi \mathit{d}\rho \wedge  \mathit{d}\xi \Big)\nn \\&~&+\frac{2\sin ^2\theta_3 \cos \theta _1}{ \alpha '^2 l_p^2}\Big(2 A\rho  \text{R}^6 t_3^6  \sin \chi  \mathit{d}\chi- \cos \chi \mathit{d}\rho \left(2 A \text{R}^6 t_3^6+m_2^2\rho ^2 \alpha '^2 l_p^2\right)\Big)\wedge d\phi_1\bigg],\nn \\
e^{-2\hat{\Phi}}&=&\frac{2\sqrt{2}l_p^3}{\text{R}^3t_3^3A^{3/2}}\Delta, \\ \Delta&=&\frac{\sqrt{2}t_3^3\text{R}^3}{\sqrt{A}l_p\alpha'^3}\big(4t_3^6\text{R}^6A\sin^2\theta_3+l_p^2m_2^2\alpha'^2\rho^2(2\cos^2\chi\sin^2\theta_3+A\sin^2\chi)\big),\nn
\eea
and $\Xi=2 A \text{R}^6 t_3^6+m_2^2 \rho ^2 \alpha '^2 l_p^2 \cos^2\chi $.
   The RR sector contains the following fluxes,
   \bea
   \hat{F}_1&=&\frac{2 m_2 l_p \left(A \cos \theta _3 \cos\chi  \mathit{d}\rho -\rho  \left(\cos \chi  \left(A\sin \theta _3+\cos\theta _3A'\right)\mathit{d}\theta _3 +A  \cos\theta _3 \sin \chi \mathit{d}\chi \right)\right)}{A^2 \sqrt{\alpha '}},\nn \\
     \hat{F}_3&=&\frac{2}{\alpha'^{3/2} l_p} \left(\frac{m_2^2 \rho \alpha'^2 \cos \theta _3 l_p^2 \mathit{d}\rho}{A}-6 \text{R}^6 t_3^6 \sin \theta _3\mathit{d}\theta_3\right)\wedge d\Omega_2^{(1)}\nn \\&~& +\frac{2 \sqrt{2} m_2^4 \rho ^3 \text{R}^3 t_3^3 \sin ^2\theta_3 \cos \theta _3 \sin 2 \chi }{A^{3/2} \Delta  \sqrt{\alpha '}} \left(\mathit{d}\xi -\cos \theta_1 \mathit{d}\phi _1\right)\wedge \mathit{d}\rho \wedge \mathit{d}\chi\nn \\&~&
   -  \frac{2 \sqrt{2} m_2^2 \rho  \text{R}^3 t_3^3 \sin \chi  \left(A \sin\theta _3+\cos \theta _3A'\right)}{A^{3/2} \Delta  \alpha '^{5/2} l_p^2} ( \mathit{d}\xi+\cos\theta_1 d\phi_1) \wedge \mathit{d}\theta _3\wedge\nn \\&~& \bigg(4 \rho  \text{R}^6 t_3^6 \sin ^2\theta _3 \cos \chi \mathit{d}\chi  +\sin \chi\left(m_2^2 \rho ^2 \alpha'^2 l_p^2+4 \text{R}^6 t_3^6 \sin ^2\theta_3\right)\mathit{d}\rho\bigg),\nn \\
       \hat{F}_5&=&8t_3^3\Big[\frac{4  \text{R}^5 t_3^6 \sin \theta _3 \cos\theta _3\mathit{d}\theta _3}{\alpha'^{3/2} l_p^2}-\frac{3 m_2^2 \rho  \sqrt{\alpha'}  \mathit{d}\rho}{\text{R}}\Big]\wedge d\Omega_{AdS_4}\nn \\&~&
 -\frac{4 \sqrt{2} m_2^3 \rho ^2 \text{R}^9 t_3^9  \sin \theta _3 \sin \chi }{\sqrt{A} \Delta  \alpha '^{3/2} l_p^2}\bigg[ 3 \rho  \left(A \sin^2\chi +2 \sin ^2\theta _3\cos ^2\chi \right)\mathit{d}\chi \wedge \mathit{d}\theta _3\nn \\&~&-3 \sin 2 \chi \mathit{d}\rho \wedge \mathit{d}\theta _3-2 \sin \theta _3 \cos \theta _3 \mathit{d}\rho \wedge \mathit{d}\chi\bigg]\wedge d\xi\wedge d\Omega_2^{(1)}
   \eea
   A simple check of the zero's of $\Delta$ for this background reveals that $\theta_3=0$ with $\rho=0$ or $\chi=0$ are the singular points (generated by the NATD).  We have also checked this by examining the Einstein frame Ricci scalar.
\section{Backgrounds from $AdS_4\times N(1,1)$ with less supersymmetry } \label{Sec:N11}
 The $N(k,l)$ spaces are $SU(3)/U(1)$ cosets where the $U(1)$ acts as \newline $diag(\exp(ik\theta),\exp(il\theta),\exp(-i(k+l)\theta))$. The particular case of $N(1,1)$ is supersymmetric and the metric can be viewed as an $SO(3)$ bundle over $\mathbb{CP}^2$,
\bea
ds^2&=&d\mu^2+\frac{1}{4}\sin^2\mu(\sigma_1^2+\sigma_2^2+\cos^2\mu\sigma_3^2)  \nonumber \\
&+&\lambda^2\big[(\Sigma_1-\cos\mu\sigma_1)^2+(\Sigma_2-\cos\mu\sigma_2)^2+(\Sigma_3-\frac{1}{2}(1+\cos^2\mu)\sigma_3)^2\big],
\eea
with $\mathcal{R}=-\frac{27}{5L^2}$.  There are two values of $\lambda$ yielding supersymmetric solutions, $\lambda^2=1/2$ corresponds to ${\cal N}=3$ and $\lambda^2=1/10$ corresponds to ${\cal N}=1$ susy. In this section we will focus on the more  supersymmetric version of the background but an analogous treatment can be applied to the less supersymmetric case. We expect a similar situation in the case of the squashed $S^7$ which also preserves less supersymmetry compared to the round sphere background. We will not pursue this direction here but it is certainly an interesting one.

In this section we consider the metric on $AdS_4\times N(1,1)$ with $\lambda^2=\frac{1}{2}$, which corresponds to the case of $\mathcal{N}=3$ supersymmetry.  There are several U(1) angles with which one can perform the reduction to IIA.  We have examined all of the possible reductions. The one that leads to a tractable solution is the U(1) isometry given by the linear combination of $\psi_1+\psi_2$.  We label $\alpha$ to be the coordinate corresponding to $\psi_1-\psi_2$.

The reduced background is given by,
\bea
\label{N11IIA}
ds^2&=&e^{\frac{2}{3}\Phi}\bigg[ \frac{1}{2}ds^2(AdS_4)+\text{R}^2 \big(d\mu^2+\frac{1}{4}\sin^2\mu ds^2(\Omega_2^{(2)})+\nn \\&~&+ \frac{1}{2}\left((s_2-\cos\mu d\theta_2)^2+(s_1-\cos\mu\sin\theta_2 d\phi_2)^2\right)+ \frac{2}{Z}\cos\mu(s_3-\cos\theta_2 d\phi_2)^2\bigg], \nn \\
B_2&=&0,\ \ e^{2\Phi}=\frac{\text{R}^3\sin^3\mu Z^{3/2}}{512 l_p^3}
\eea
in which we have recovered an SU(2) isometry in ($\theta_1,\phi_1,\alpha$), characterised by the Maurer-Cartan forms $s_i$.  We also define $Z=3+\cos2\mu$.  The RR fluxes are given by,
\bea
C_1&=&\frac{l_p}{Z}((-5+\cos2\mu)d\alpha-8\cos\theta_1d\phi_1+4\cos\theta_2\sin^2\mu d\phi_2), \ F_2=dC_1\nn \\
F_4&=&-\frac{3}{2\sqrt{2}\text{R}}d\Omega_{AdS_4}.
\eea

\subsection{NATD($\Downarrow_{\beta}(AdS_4\times N(1,1))$)}
In this section, we present the results of a Non-Abelian T-duality applied to the SU(2) isometry in Eq. (\ref{N11IIA}) above.  This background is somewhat unique compared to the other backgrounds presented in this paper and other NATD backgrounds presented thus far in the literature.  It is unique in the sense that there is mixing between the spectator coordinates ($\theta_2,\phi_2$) with the Maurer Cartan forms, $s_1,s_2$.  Mixing terms with $s_3$ have been fairly common (ex. the ABJM, Klebanov Witten backgrounds), however, the non-symmetric mixing with $s_1,s_2$ leads to a rich Type IIB NATD background.  An interesting consequence of this mixing is the breaking of the U(1) isometry normally found in the $\xi$ coordinate after NATD.
\bea
\hat{ds}^2&=& e^{\frac{2}{3}\Phi}\big[\frac{1}{2}ds^2(AdS_4)+\text{R}^2 \big(d\mu^2+\frac{1}{4}\sin^2\mu ds^2(\Omega_2^{(2)})\big)\big]+\hat{ds}^2_3,\nn \\
\hat{ds}^2_3&=&\frac{m_2^2 \text{R}^6 \sin ^2\mu}{64\Delta\alpha'l_p^2} \Big[ \cos ^2\mu \rho ^2  \left(\cos \mu \sin \xi  \cos \chi\mathit{d}\theta _2 -\text{G}\mathit{d}\phi _2+ \sin \chi\mathit{d}\xi \right){}^2\nn \\&~&+\frac{1}{4K} \left(\rho  \text{K}\left(\cos \mu  \left(\sin \theta _2 \sin \xi\mathit{d}\phi _2-\cos \xi \mathit{d}\theta _2\right)+\mathit{d}\chi \right)-2 \sin ^2\mu  \sin \chi  \cos \chi \mathit{d}\rho \right){}^2\Big]\nn \\&~&+\frac{16 m_2^2 \sqrt{Z} \alpha '^2  l_p \csc \mu}{\text{R}^3 \text{K}}\mathit{d}\rho ^2,\nn \\
\hat{B}_2&=&\frac{m_2 \text{R}^3 \sin \mu }{1024 \Delta  \sqrt{Z} \alpha '^2l_p^3}\bigg[\big(\text{H} \rho  \text{R}^6 Z \sin ^2\mu\cos ^3\mu  \mathit{d}\theta _2-\text{G} \rho  \text{R}^6 Z \sin ^2\mu  \cos ^2\mu \mathit{d}\chi \nn \\&~&+\text{J}\text{R}^6 Z \sin ^2\mu  \cos ^2\mu \mathit{d}\rho \big)\wedge \mathit{d}\phi _2+64 m_2^2 \rho^2\alpha '^2 l_p^2  \bigg( K  d\hat{\Omega}_2\nn \\&~&-\big(\text{M} \cos \mu  \mathit{d}\phi _2+\sin ^2\mu  \sin \chi  \sin 2 \chi  \mathit{d}\xi  +Z  \cos \mu  \sin \xi \sin \chi   \mathit{d}\theta _2\big)\wedge d\rho \bigg)\nn \\&~&+ \text{R}^6 Z\cos^3\mu \Big(\rho  \sin ^2\mu  \cos \xi  \sin \chi  \mathit{d}\xi \wedge \mathit{d}\theta _2+ \sin ^2\mu   \sin \xi \sin \chi  \mathit{d}\rho\wedge \mathit{d}\theta _2\nn \\&~&+\rho  \sin ^2\mu\sin\xi  \cos \chi  \mathit{d}\chi \wedge \mathit{d}\theta _2-\rho \sin \theta _2 \sin ^2\mu    \sin \xi  \sin \chi  \mathit{d}\xi\wedge \mathit{d}\phi _2\Big)\bigg],\nn \\
e^{-2\hat{\Phi}}&=&\frac{512l_p^3}{\text{R}^3\sin^3\mu Z^{3/2}}\Delta, \\ \Delta&=&\frac{\text{R}^3\sin\mu}{1024l_p^3\sqrt{Z}}\Big(\text{R}^6\cos^2\mu\sin^2\mu Z+64l_p^2\alpha'^2m_2^2\rho^2(4\cos^2\mu\cos^2\chi+\sin^2\chi Z)\Big).\nn
\eea
To make the presentation of the background slightly more succinct we have had to define a number of functions,
\bea
K&=&4 \cos ^2\mu  \cos ^2\chi +Z \sin ^2\chi,
\nn \\ G&=&\cos \theta _2\sin \chi -\sin \theta _2 \cos \mu  \cos \xi  \cos\chi,\nn \\
H&=&\cos \theta _2 \cos \xi  \sin \chi -\sin\theta _2 \cos \mu  \cos\chi,\nn \\
J&=&\sin \theta _2\cos \mu  \cos \xi  \sin \chi +\cos \theta _2 \cos\chi,\nn \\ M&=&4 \cos \theta _2\cos \mu \cos \chi +Z \sin \theta _2 \cos \xi  \sin \chi, \nn\\ N&=&\sin ^2\mu  \sin ^2\chi +2 \cos ^2\mu,\nn \\ V&=&64 m_2^2 \rho ^2 \alpha '^2 l_p^2 \sin ^2\chi +\text{R}^6 \sin ^2\mu  \cos ^2\mu,\nn \\  X&=&\text{R}^9 \sin ^5\mu  \cos ^2\mu  \cos ^2\chi -512 \Delta  \sqrt{Z}\alpha '^3 l_p^3.
\eea
The reduction to Type IIA along $\psi_1+\psi_2$ generates a singularity at $\mu=0$, while the NATD adds singular points at $\mu=\frac{\pi}{2}$ with $\rho=0$ or $\chi=0$ to the dual background.

The dual RR sector can be written as,
\bea
\hat{F}_1&=&\frac{8 m_2 l_p}{K Z^2 \sqrt{\alpha '}} \left(K \rho  Z  \sin \chi\mathit{d}\chi  -2 \cos \chi \left(2 K \rho  \sin \mu  \cos \mu\mathit{d}\mu \right.\right.\nn \\&~&\left.\left.+Z \left(\sin ^2\mu  \sin ^2\chi+2 \cos ^2\mu \right)\mathit{d}\rho  \right)\right),\nn \\
\hat{F}_3&=&\frac{\text{R}^9 m_2^2\rho ^2 \cos ^3\mu \sin ^4\mu   \cos \chi  }{32 Z^{3/2} \Delta  l_p^2 \alpha  '^{5/2}} \bigg[ \mathit{d}\mu \wedge \mathit{d}\chi \wedge \big( \text{G} \mathit{d}\phi _2- \cos\mu  \cos \chi   \sin \xi \mathit{d}\theta _2\big)\nn \\&~&+ \sin \chi  \mathit{d}\mu \wedge  \mathit{d}\xi \wedge\Big( \mathit{d}\chi
- \cos \mu  \cos \xi \mathit{d}\theta _2
+ \cos \mu   \sin \xi  \sin \theta _2 \mathit{d}\phi _2\Big)\bigg]\nn \\&~&
+\frac{m_2^2 \text{N}\rho  \cos ^3\mu \sin ^2\mu  \sin\chi    \cos \chi \text{R}^9}{64 \sqrt{Z} \Delta  \text{K}l_p^2 \alpha '^{5/2}} \mathit{d}\xi \wedge \mathit{d}\rho \wedge\Big( \sin \mu \cos \xi \mathit{d}\theta _2-\sin \xi \sin \theta _2  \mathit{d}\phi _2\Big)\nn \\&~&+\frac{m_2^2\text{R}^9\rho ^2 \cos ^3\mu \sin ^3\mu  \sin ^2\chi}{128 \sqrt{Z} \Delta  l_p^2 \alpha '^{5/2}} \mathit{d}\xi\wedge \mathit{d}\chi \wedge\Big( -\cos \xi  \mathit{d}\theta _2  +  \sin \xi  \sin \theta _2 \mathit{d}\phi _2 \Big)\nn \\&~&+\frac{m_2^2\text{R}^3\rho \cos \mu \sin \mu }{128 \sqrt{Z} \Delta  l_p^2 \alpha '^{5/2}} \mathit{d}\chi \wedge \mathit{d}\theta _2\wedge\Big(\text{R}^6\text{H} \rho  \cos ^2\mu  \sin ^2\mu  \sin\chi   \mathit{d}\phi _2 -\text{V}   \sin \xi   \mathit{d}\rho \Big) \nn \\&~&
-\frac{\text{R}^3 \cos\mu \sin \mu }{128 Z^{3/2} \Delta  l_p^2 \alpha '^{5/2}}\bigg[4\text{R}^3\sin ^2\mu  \left(\text{H} \text{R}^3 \rho ^2 \cos \chi  \sin \mu  m_2^2 \cos ^3\mu \right.\nn \\&~&\left.+3 Z^{3/2} \Delta  \sin \theta _2 l_p \alpha'\right)\mathit{d}\mu \wedge \mathit{d}\theta_2\wedge \mathit{d}\phi _2 \nn \\&~&
-\rho  m_2^2 \left(256 \rho ^2 \cos \mu  \cos \chi  \cos \theta _2\sin \chi l_p^2 m_2^2 \alpha '^2+\text{V} Z \cos \xi \sin \theta _2\right)\mathit{d}\rho \wedge \mathit{d}\chi \wedge \mathit{d}\phi_2\bigg] \nn \\&~&+\frac{2 \rho ^3 \cos ^2\mu\cos \chi  \sin \mu  \sin \chi  m_2^4 \mathit{d}\xi \wedge \mathit{d}\rho \wedge\mathit{d}\chi \text{R}^3}{Z^{3/2} \Delta  \sqrt{\alpha '}}\nn \\&~&-\frac{\rho  m_2^2 }{64 Z^{3/2} \Delta  \text{K} l_p^2 \alpha '^{5/2}} \left(\frac{1}{4} \text{R}^9 Z \cos ^2\mu  \sin ^3\mu   \left(-Z \sin ^2\mu  \sin ^2\chi \sin \theta _2\right.\right.\nn \\&~&\left.4 \cos \mu  \cos \chi  \left(\text{H} \text{N}-\cos \mu  \cos \chi  \sin ^2\mu \sin \theta _2\right)\right)\nn \\&~&\left.-16 \text{R}^3 \rho ^2 \text{K}^2 \sin ^3\mu \sin \theta_2 l_p^2 m_2^2 \alpha '^2\right)\mathit{d}\rho \wedge \mathit{d}\theta _2\wedge \mathit{d}\phi _2\nn \\&~&+\frac{m_2^2\text{X} \rho  \sin \chi }{16 Z^{3/2} \Delta  \text{K}l_p^2 \alpha'^{5/2}}\mathit{d}\mu \wedge \mathit{d}\rho \wedge \Big( \cos ^2\mu \cos \chi  \sin \mu  \sin  \xi   \mathit{d}\theta_2-  \text{G}  \mathit{d}\phi _2+\sin\chi d\xi\Big),\nn \\
\hat{F}_5&=&\frac{m_2^3\rho^2\text{R}^9\sin^3\mu\cos\mu}{256\Delta\sqrt{Z}\alpha'^{3/2}l_p^2}\bigg[\big(-\frac{3}{2} K   \sin\mu    \mathit{d}\mu+\cos \mu \left(\sin ^2\mu -2 \cos ^2\mu \right)  \mathit{d}\rho\big) \wedge d\hat{\Omega}_2\nn \\&~&+3  \sin^3\mu   \sin ^2\chi  \cos \chi  \mathit{d}\mu \wedge \mathit{d}\xi \wedge \mathit{d}\rho\bigg]\wedge d\Omega_2^{(2)}\nn \\&~&+\big(\frac{3 m_2^2 \rho  \sqrt{\alpha '}  \mathit{d}\rho }{2 \sqrt{2} \text{R}}-\frac{\text{R}^5 \sin \mu  \cos \mu \left(\sin ^2\mu -2 \cos ^2\mu \right)}{128 \sqrt{2} \alpha '^{3/2} l_p^2}\mathit{d}\mu\big)\wedge d\Omega_{AdS_4}.
   \eea
\section{Dual CFT Central Charge}
\label{Sec:CC}
As a first step into the interpretation of the new backgrounds we produce in this manuscript we perform an analysis of the central charge of the dual field theories.
In order to compute the dual field theory central charge, we first consider the quantized Page charges \cite{Page:1984qv,Marolf:2000cb}, defined by
\bea
\label{QPage}
Q_{M2}&=&\frac{1}{2\kappa_{11}^2T_{M2}}\int_{\Sigma_{7}}\star F_{4}=M2,\\
Q_{Dp}&=&\frac{1}{2\kappa_{10}^2T_{Dp}}\int_{\Sigma_{8-p}}\big(\sum_iF_i\big)\wedge e^{-B_2}=N_{Dp},
\eea
where $\kappa_{11}^2=(2\pi)^8l_p^9,\ \kappa_{10}^2=(2\pi)^7\alpha'^4$ and the brane tensions are $T_{M2}=\frac{1}{(2\pi)^2l_p^3}$, and  $T_{Dp}=\frac{1}{(2\pi)^p\alpha'^{\frac{p+1}{2}}}$.
We will present the general results of the Page charges and central charge for a few sample backgrounds considered in this manuscript.  In particular, we discuss the importance of the factors of the 11d radius R, and their role in the N scaling of the central charge.  Our method for computing the central charge is based on \cite{Klebanov:2007ws}, as adjusted and generalized in \cite{Macpherson:2014eza}. The main modifications take into consideration a potential dependence on the coordinates of the internal manifold perpendicular to the field theory directions.
A generic string frame metric in type II string theory, dual to a QFT in ($d+1$)-dimensions is defined to have the following form,
\beq
ds^2= a dz_{1,d}^2 + ab dr^2 +\text{R}^2g_{ij}d\theta^i d\theta^j.
\label{metricvvv}\eeq
As in \cite{Macpherson:2014eza}, we define the modified internal volume to be,
\beq
\hat{V}_{int}=\int d\vec{\theta} \sqrt{e^{-4\Phi}\det[g_{int}] a^d},
\label{hatV}\eeq
so that the function $\hat{H}$ is in general given by,
\beq
\hat{H}= \hat{V}_{int}^2.
\eeq
Then, the central charge for a QFT in $(d+1)$ spacetime dimensions is
defined to be
\cite{Klebanov:2007ws}:
\beq
\label{Eq:CC}
c= d^d \frac{b^{d/2} \hat{H}^{(2d+1)/2}}{G_N(\hat{H}')^d}.
\eeq
The $G_N$ factor is needed to cancel the length dimensions in $\hat{H}$.  We will use $G_{11}=l_p^9$, and $G_{10}=l_s^8=\alpha'^4$.
In our case of $AdS_4$, the functions are,
\beq
a=\frac{r^2}{R^2}f(\theta^i),\;\;\;b=\frac{R^4}{r^4},\;\;\; d=2.
\eeq
In the case of 11 spacetime dimensions, we obtain $\hat{V}_{int}\sim \text{R}^9$, $\hat{H}\sim \text{R}^{18}r^3$, therefore, $c\sim\frac{L^9}{l_p^9}$.  Due to our choice of Ansatz, Eq. (\ref{Freund-Rubin}), the relevant objects are always M2 branes, whose normalization yields a scaling relation, $\text{R}^6\sim l_p^6 N_{M2}$, which means $c\sim N_{M2}^{3/2}$, as known previously in \cite{Aharony:1999ti}, for example.  In all of the reductions to Type IIA considered in this paper the $F_4$ flux comes through the reduction unaffected and an $F_2=dC_{(1)}$ flux is generated.  The normalization of the $D2$ branes leads to a relation, $\text{R}^6\sim l_p\alpha'^{5/2}N_{D2}$.  The `raw' central charge after the reduction now scales like, $c_{IIA}\sim \frac{\text{R}^9}{l_p\alpha'^4}$, therefore $c_{IIA}\sim \frac{\sqrt{l_p}}{\alpha'^{1/4}}N_{D2}^{3/2}$.  We also note the presence of D6 branes in the Type IIA backgrounds, however they are independent of R and   instead are scaled with factors of $l_p$ only, suggesting they are topological charges.  In the case of ABJM, the factor of k present in $F_2$ has the field theory interpretation as the level number of the dual gauge theory.


The relevant objects to the central charge after the NATD are the $D5$ branes, as they always inherit a term with the $\text{R}^6$ scaling factor.
As observed previously in \cite{Lozano:2014ata} for the ABJM background, we see that the effect of NATD is to preserve the $N^{3/2}$ scaling, but to change the numerical coefficient of the central charge.

After the NATD we wish to constrain the range of the dual coordinate $\rho$, so that we may compute the internal Volume, $\hat{V}_{int}$.  Following the prescription first hinted at in \cite{Lozano:2013oma}, and further discussed in \cite{Lozano:2014ata,Macpherson:2014eza}, we compute the periodic quantity $b_0$ defined by,
\be
\label{b0def}
b_0=\frac{1}{4\pi^2\alpha'}\oint_{\Sigma_2}\hat{B}_2\subset [0,1].
\ee
The relevant two-cycle is,
\be
\Sigma_2=[\chi,\xi].
\ee
In each case we restrict to a submanifold, which we specify below:
\begin{center}
\begin{tabular}{ c  c }
Background & Submanifold\\
NATD($\Downarrow_{\tau}(AdS_4\times S^7/\mathbb{Z}_k)$)  & $\alpha=0$, $\rho$ fixed\\
NATD($\Downarrow_{\psi_2}(AdS_4\times S^7)$) & $\mu=0$ \\
NATD($\Downarrow_{\tau}(AdS_4\times M^{1,1,1})$) & $\mu=0$\\
NATD($\Downarrow_{\phi_3}(AdS_4\times Q^{1,1,1}))$ & $\theta_3=0$, $\rho$ fixed\\
NATD($\Downarrow_{\beta}(AdS_4\times N(1,1)$) & $\mu=0$, $\rho$ fixed
\end{tabular}
\end{center}
We should point out that in all of these cases the cycle is placed on a singularity, which could be problematic.Ê However, the strategy still leads to reasonable results and was also observed in the case of NATD of $AdS_5\times S^5$ in \cite{Macpherson:2014eza}.Ê Presumably in a full homology theory, it would be possible to show independence of this procedure on position in the manifold, but we leave that question to future investigations.
For the remainder of this section we assume the prescription can be trusted, and move on to compute,
\be
b_0=\frac{1}{4\pi^2\alpha'}\oint_{\Sigma_2}(\alpha'\rho\sin\chi)=\frac{\rho}{\pi}\subset [0,1],
\ee
where $0\leq \chi \leq \pi$, $0\leq \xi \leq 2\pi$, and we have absorbed the $m_2$ gauge fixing constant into the definition of $\rho$.
However, as discussed in \cite{Macpherson:2015tka}, there is one NS-five brane every
time $\rho$ crosses integer multiples of $\pi$. Therefore we take the range of $\rho$ from  $0\leq \rho \leq (n+1)\pi$.
Next we compute $H_3=d\hat{B}$,
\be
H_3=\alpha' \sin\chi  d\rho\wedge d\xi\wedge d\chi.
\ee
Following \cite{Macpherson:2015tka}, we normalize the NS flux to $N_{NS5}$ using $T_{NS5}=\frac{1}{(2\pi)^5}$,
\be
Q_{NS5}=\frac{1}{2\kappa_{10}^2T_{NS5}}\int H_3= (n+1)=N_{NS5}
\ee
We are now in a position to compute the Page charges and dual field theory central charge after the NATD.  The results are summarized in Tables \ref{ABJMtable}, \ref{S7table}, \ref{M111table}, \ref{Q111table}, and \ref{N11table} below for the backgrounds with $S^7/\mathbb{Z}_k\footnote{A careful reader will notice a factor of 2 difference between our results for the central charge and what was found in \cite{Lozano:2014ata} for the ABJM background.  This is due to the Jacobian factor when changing coordinates from $\alpha\to 2\zeta$.},S^7,M^{1,1,1},Q^{1,1,1}$, and $N(1,1)$, respectively.  In the first column we present the flux normalization result, obtained from Eq. (\ref{QPage}) and the corresponding cycle used to integrate.  The second column contains the `Raw' central charge, which is directly computed from Eq. (\ref{Eq:CC}).  The third and final column is the substitution of the flux normalization into the `Raw' central charge.  In the cases where we have left some of the numerical constants generalized, we also present the result with the values of these constants replaced.

As noted in \cite{Lozano:2014ata}, NATD maps integer charges onto non-integer charges, due to a violation of the condition, $T_{p-n}=(2\pi)^nT_p$ on the D brane tensions.  In fact, we see a generic difference of $\frac{\pi}{2}$ between the D2 and D5 brane normalizations in these cases.  In the case of $AdS_4\times S^7/\mathbb{Z}_k$, we note, as in \cite{Lozano:2014ata}, that after the NATD, $k$ is no longer a well-defined level.  We define a new level for the dual theory according to,
\be
k_5=\frac{1}{2\kappa_{10}^2T_{D5}}\int_{\Sigma_3}(\hat{F}_3-\hat{B}_2\wedge \hat{F}_1)=\frac{l_p}{\sqrt{\alpha'}}\frac{m_2^2\pi^3}{2}k N_{NS5}^2,
\ee
where the cycle of integration is $\Sigma_3=[\rho,\theta_1,\phi_1]$, and $\rho$ is integrated over $[0, (n+1)\pi]$.  This leads to a $N_{NS5}^2$ scaling\footnote{We are thankful to Niall Machpherson for pointing this out to us.} in the central charge, which differs from the other cases.  In the cases without $k$, a factor of $N_{NS5}^3$ arises from integration over $\rho^2$ in $\hat{V}_{int}$.

Finally, we consider changes in the Page charges under large gauge transformations in $B_2$, particularly $\Delta B_2=n\pi\alpha' \sin\chi d\chi\wedge d\xi$.  The Page charge associated to the $D5$ branes is always either zero or independent of \text{R} on the cycle of interest.  Therefore, we compute $\Delta Q_{D3}$ using
\be
\Delta Q_{D3}=-\frac{1}{2\kappa_{10}^2 T_{D3}}\int_{\Sigma_5}(-\Delta B_2\wedge \hat{F}_3+\frac{1}{2}\Delta B_2\wedge \Delta B_2\wedge \hat{F}_1),
\ee
and $\Sigma_5=\Sigma_3+[\chi,\xi]$.  In all of the cases examined, we find the relation,
\be
\Delta Q_{D3}=n Q_{D5}.
\ee

 \subsection{Dual CFT Central Charge for Coset NATD}
 In this section, we present a proposal to investigate the effect of large gauge transformations on cases where the NATD is performed on an $SU(2)/U(1)$ coset isometry.  The same method used above in Eq. (\ref{b0def}) cannot be directly applied here, as there is no 2-cycle with which to compute $b_0$ and thus, restrict the range of the dual coordinates.   We propose to introduce a nonzero $B_2$ which is closed, and therefore does not affect the equations of motion:  $\tilde{B}_2=\alpha' d\rho\wedge d\chi$.  This is equivalent to a large gauge transformation on $B_2$.  We compute $b_0$,
\be
b_0=\frac{1}{4\pi^2\alpha'}\int_{\rho_o\chi_o} \tilde{B}_2=\frac{\rho_o\chi_o}{2},
\ee
where $\rho$ and $\chi$ are integrated over $[0,\rho_o\pi]$ and $[-\chi_o\pi,\chi_o\pi]$, respectively.  Demanding $b_0= [0,1]$ restricts the range of $\rho$ and $\chi$.  Normally for polar coordinates, we would expect the $\chi$ angle to take the range $0\leq\chi\leq 2\pi$ (or $-\frac{\pi}{2}\leq\chi\leq\frac{\pi}{2}$), however the modified internal volume $\hat{V}_{int}$ computed from Eq. (\ref{hatV}), restricts the range of $\chi$ to be $-\frac{\pi}{2}\leq\chi\leq \frac{\pi}{2}$, due to a factor of $\cos\chi$.  Then we see that $\rho$ would have to take the range $0\leq\rho\leq4\pi$.
Note that this condition was different in the cases of SU(2) NATD.

Considering the case of the $\tau$ reduction of $AdS_4\times Q^{1,1,1}$ as an example, we compute the Page charge for D3 branes,
\be
\hat{F}_5-\tilde{B}_2\wedge \hat{F}_3=-\frac{3m_2t_2^6t_3}{t_1}\frac{\text{R}^6}{l_p\sqrt{\alpha'}}(\sin\chi d\rho+\rho\cos\chi d\chi)\wedge d\Omega_2^{(1)}\wedge d\Omega_2^{(2)}.
\ee
Integrating over the cycle containing $\rho$ and the two 2-spheres, we obtain the normalization condition,
\be
\frac{1}{2\kappa_{10}^2T_{D3}}\int \hat{F}_5-\tilde{B}_2\wedge \hat{F}_3=\frac{12m_2t_2^6t_3}{t_1\pi}\frac{\text{R}^6}{l_p\alpha'^{5/2}}=N_{D3}.
\ee
After the NATD, we find that the central charge is,
\bea
c&=&\frac{512m_2^3t_1^2t_2^6t_3\pi^5}{3}\frac{\text{R}^9}{l_p\alpha'^4}=\frac{64m_2^{3/2}t_1^{7/2}\pi^{13/2}}{9\sqrt{3}t_2^3\sqrt{t_3}}\frac{\sqrt{l_p}}{\alpha'^{1/4}}N_{D3}^{3/2}\nn \\&=&\frac{256m_2^{3/2}\pi^{13/2}}{9\sqrt{3}}\frac{\sqrt{l_p}}{\alpha'^{1/4}}N_{D3}^{3/2}.
\eea
If we had instead integrated over the $\chi$ plus the two 2-spheres cycle, the final central charge would be different by a factor of $\sqrt{2}$.


\begin{landscape}
\begin{table}
\caption{Results for $AdS_4\times S^7/\mathbb{Z}_k$ and its supergravity duals.}
\label{ABJMtable}
\begin{center}
\begin{tabular}{| c || c | c | c |}
\hline & & & \\
Background &  \begin{tabular}{c}Page Charge \\[4pt] Cycle\end{tabular}  & `Raw' CC & CC  \\ & & &  \\
\hline & &  & \\
$AdS_4\times S^7/\mathbb{Z}_k$ & \begin{tabular}{c} $Q_{M2}=\left(\frac{\text{R}}{l_p}\right)^6\frac{2\lambda_2^6\lambda_3}{k^2\pi^2}=N_{M2} $\\[4pt]$\Sigma_7=[\alpha,\theta_1,\phi_1,\theta_2,\phi_2,\psi,\tau]$\end{tabular} & $\frac{4\lambda_2^6\lambda_3\pi^4}{3k^2}\left(\frac{\text{R}}{l_p}\right)^9$ &\begin{tabular}{c} $\frac{\sqrt{2}\pi^7k}{3\lambda_2^3\sqrt{\lambda_3}}N_{M2}^{3/2}$ \\[4pt] $=\frac{16}{3}\sqrt{2}\pi^7kN_{M2}^{3/2}$\end{tabular} \\ & & & \\
\hline & & & \\
$\Downarrow_{\tau}(AdS_4\times S^7/\mathbb{Z}_k)$ & \begin{tabular}{c}$Q_{D2}=\frac{\text{R}^6}{l_p\alpha'^{5/2}}\frac{2\lambda_2^6\lambda_3}{k\pi^2}=N_{D2}$ \\[4pt] $\Sigma_6=[\alpha,\theta_1,\phi_1,\theta_2,\phi_2,\psi]$\end{tabular} & $\frac{2\lambda_2^6\lambda_3\pi^3}{3k}\frac{\text{R}^9}{l_p\alpha'^4}$ &\begin{tabular}{c} $\frac{\pi^6\sqrt{k}}{3\sqrt{2}\lambda_2^3\sqrt{\lambda_3}}\frac{\sqrt{l_p}}{\alpha'^{1/4}}N_{D2}^{3/2}$ \\[4pt] $=\frac{8}{3}\sqrt{2}\pi^6\sqrt{k}\frac{\sqrt{l_p}}{\alpha'^{1/4}}N_{D2}^{3/2}$\end{tabular} \\ & & &  \\
\hline & & & \\
NATD($\Downarrow_{\tau}(AdS_4\times S^7/\mathbb{Z}_k)$) & \begin{tabular}{c}$Q_{D5}=\frac{\text{R}^6}{l_p\alpha'^{5/2}}\frac{\lambda_2^6\lambda_3}{k\pi}=N_{D5}$ \\[4pt]$\Sigma_3=[\alpha,\theta_1,\phi_1]$\end{tabular} & $\frac{m_2^3\pi^5\lambda_2^6\lambda_3}{18k}\frac{\text{R}^9}{l_p\alpha'^4}$ & \begin{tabular}{c}$\frac{m_2^2\pi^6\sqrt{k_5}}{9\sqrt{2}\lambda_2^3\sqrt{\lambda_3}}N_{NS5}^2N_{D5}^{3/2}$\\[4pt] $=\frac{8\sqrt{2}m_2^2\pi^6\sqrt{k_5}}{9}N_{NS5}^2N_{D5}^{3/2}$\end{tabular} \\ & & & \\
\hline
\end{tabular}
\end{center}
\end{table}
\begin{table}
\caption{Results for $AdS_4\times S^7$ and its supergravity duals.}
\label{S7table}
\begin{center}
\begin{tabular}{| c || c | c | c |}
\hline & & & \\
Background & \begin{tabular}{c}Page Charge\\[4pt] Cycle\end{tabular} & `Raw' CC & CC  \\ & & &  \\
\hline & &  & \\
$AdS_4\times S^7$ & \begin{tabular}{c}$Q_{M2}=\left(\frac{\text{R}}{l_p}\right)^6\frac{2}{\pi^2}=N_{M2}$\\[4pt] $\Sigma_7=[\mu,\theta_1,\phi_1,\psi_1,\theta_2,\phi_2,\psi_2]$\end{tabular} & $\frac{32\pi^4\text{R}^9}{3l_p^9}$ & $\frac{8\sqrt{2}}{3}\pi^7N_{M2}^{3/2}$  \\ & & & \\
\hline & & & \\
$\Downarrow_{\psi_2}(AdS_4\times S^7)$ & \begin{tabular}{c}$Q_{D2}=\frac{\text{R}^6}{l_p\alpha'^{5/2}}\frac{1}{\pi^2}= N_{D2}$\\[4pt]$\Sigma_6=[\mu,\theta_1,\phi_1,\psi_1,\theta_2,\phi_2]$\end{tabular} & $\frac{8\pi^3}{3}\frac{\text{R}^9}{l_p\alpha'^4}$ & $\frac{8\pi^6}{3}\frac{\sqrt{l_p}}{\alpha'^{1/4}}N_{D2}^{3/2}$ \\ & & &  \\
\hline & & & \\
NATD($\Downarrow_{\psi_2}(AdS_4\times S^7)$) & \begin{tabular}{c}$Q_{D5}=\frac{\text{R}^6}{l_p\alpha'^{5/2}}\frac{1}{2\pi}= N_{D5}$\\[4pt]$\Sigma_3=[\mu,\theta_2,\phi_2]$\end{tabular} & $\frac{2m_2^3\pi^5}{9}\frac{\text{R}^9}{l_p\alpha'^4}N_{NS5}^3$ & $\frac{4\sqrt{2}m_2^3\pi^{13/2}}{9}\frac{\sqrt{l_p}}{\alpha'^{1/4}}N_{NS5}^3N_{D5}^{3/2}$ \\ & & & \\
\hline
\end{tabular}
\end{center}
\end{table}
\begin{table}
\caption{Results for $AdS_4\times M^{1,1,1}$ and its supergravity duals.}
\label{M111table}
\begin{center}
\begin{tabular}{| c || c | c | c |}
\hline & & & \\
Background & \begin{tabular}{c}Page Charge\\ [4pt] Cycle\end{tabular} & `Raw' CC & CC \\ & & &  \\
\hline & &  & \\
$AdS_4\times M^{1,1,1}$ & \begin{tabular}{c}$Q_{M2}=\left(\frac{\text{R}}{l_p}\right)^6\frac{3c_2c_3^3c_4^4c_5^3}{c_1\pi^6}=N_{M2}$\\[4pt] \\  $\Sigma_7=[\mu,\theta_1,\phi_1,\theta_2,\phi_2,\psi,\tau]$\end{tabular} & $16c_1^2c_2c_3^2c_4^4c_5^3\pi^4\left(\frac{\text{R}}{l_p}\right)^9$ &\begin{tabular}{c} $\frac{16\pi^7c_1^{7/2}}{3\sqrt{3}\sqrt{c_2}c_3c_4^2c_5^{3/2}}N_{M2}^{3/2}$ \\[4pt] \\ $=\frac{128\pi^7}{9\sqrt{3}}N_{M2}^{3/2}$\end{tabular} \\ & & & \\
\hline & & & \\
$\Downarrow_{\tau}(AdS_4\times M^{1,1,1})$ & \begin{tabular}{c}$Q_{D2}=\frac{\text{R}^6}{l_p\alpha'^{5/2}}\frac{3c_2c_3^3c_4^4c_5^3}{2c_1\pi^2}=N_{D2}$\\[4pt] \\ $\Sigma_6=[\mu,\theta_1,\phi_1,\theta_2,\phi_2,\psi]$\end{tabular} & $4c_1^2c_2c_3^2c_4^4c_5^3\pi^3\frac{\text{R}^9}{l_p\alpha'^4}$ & \begin{tabular}{c} $\sqrt{\frac{2}{3}}\frac{8c_1^{7/2}\pi^6}{3\sqrt{c_2}c_3c_4^2c_5^{3/2}}\frac{\sqrt{l_p}}{\alpha'^{1/4}}N_{D2}^{3/2}$\\[4pt] \\ $=\sqrt{\frac{2}{3}}\frac{64\pi^6}{9}\frac{\sqrt{l_p}}{\alpha'^{1/4}}N_{D2}^{3/2}
$\end{tabular} \\ & & &  \\
\hline & & & \\
NATD($\Downarrow_{\tau}(AdS_4\times M^{1,1,1})$) & \begin{tabular}{c}$Q_{D5}=\frac{\text{R}^6}{l_p\alpha'^{5/2}}\frac{4c_1\pi}{3c_2c_3^2c_4^4c_5^3}=N_{D5}
$\\[4pt] \\ $\Sigma_3=[\mu,\theta_1,\phi_1]$\end{tabular} & $\frac{c_1^2c_2c_3^3c_4^4c_5^3\pi^5}{3}\frac{R^9}{l_p\alpha'^4}N_{NS5}^3$ & \begin{tabular}{c} $\frac{8c_1^{7/2}\pi^{13/2}}{9\sqrt{3}\sqrt{c_2}c_3c_4^4c_5^{3/2}}\frac{\sqrt{l_p}}{\alpha'^{1/4}}N_{NS5}^3 N_{D5}^{3/2}$\\[4pt] \\ $=\frac{64\pi^{13/2}}{27\sqrt{3}}\frac{\sqrt{l_p}}{\alpha'^{1/4}}N_{NS5}^3N_{D5}^{3/2}$\end{tabular} \\ & & & \\
\hline
\end{tabular}
\end{center}
\end{table}
\begin{table}
\caption{Results for $AdS_4\times Q^{1,1,1}$ and its supergravity duals.}
\label{Q111table}
\begin{center}
\begin{tabular}{| c || c | c | c |}
\hline & & & \\
Background & \begin{tabular}{c}Page Charge\\ [4pt] Cycle\end{tabular} & `Raw' CC & CC \\ & & &  \\
\hline & &  & \\
$AdS_4\times Q^{1,1,1}$ & \begin{tabular}{c}$Q_{M2}=\left(\frac{\text{R}}{l_p}\right)^6\frac{12t_2^6t_3}{\pi^2t_1}=N_{M2}
$\\[4pt] \\  $\Sigma_7=[\mu,\theta_1,\phi_1,\theta_2,\phi_2,\psi,\tau]$\end{tabular} & $64\pi^4t_1^2t_2^6t_3\left(\frac{\text{R}}{l_p}\right)^9$ &\begin{tabular}{c} $\frac{8\pi^7t_1^{7/2}}{3\sqrt{3}t_2^3\sqrt{t_3}}N_{M2}^{3/2}$ \\[4pt] \\ $=\frac{32\pi^7}{3\sqrt{3}}N_{M2}^{3/2}$\end{tabular} \\ & & & \\
\hline & & & \\
$\Downarrow_{\phi_3}(AdS_4\times Q^{1,1,1})$ & \begin{tabular}{c}$Q_{D2}=\frac{\text{R}^9}{l_p\alpha'^{5/2}}\frac{6t_2^6t_3}{\pi^2t_1}=N_{D2}$\\[4pt] \\ $\Sigma_6=[\theta_3,\theta_1,\phi_1,\theta_2,\phi_2,\psi]$\end{tabular} & $16\pi^3t_1^2t_2^6t_3\frac{\text{R}^9}{l_p\alpha'^4}$ & \begin{tabular}{c} $\sqrt{\frac{2}{3}}\frac{4\pi^6t_1^{7/2}}{3t_2^3\sqrt{t3}}\frac{\sqrt{l_p}}{\alpha'^{1/4}}N_{D2}^{3/2}$\\[4pt] \\ $=\sqrt{\frac{2}{3}}\frac{16\pi^6}{3}\frac{\sqrt{l_p}}{\alpha'^{1/4}}N_{D2}^{3/2}$\end{tabular} \\ & & &  \\
\hline & & & \\
NATD($\Downarrow_{\phi_3}(AdS_4\times Q^{1,1,1})$) & \begin{tabular}{c}$Q_{D5}=\frac{\text{R}^6}{l_p\alpha'^{5/2}}\frac{24t_3^6}{\pi}=N_{D5}
$\\[4pt] \\ $\Sigma_3=[\theta_3,\theta_1,\phi_1]$\end{tabular} & $\frac{256m_2^3t_3^9\pi^5}{3}\frac{R^9}{l_p\alpha'^4}N_{NS5}^3$ & $\sqrt{\frac{2}{3}}\frac{8\pi^{13/2}m_2^3}{9}\frac{\sqrt{l_p}}{\alpha'^{1/4}}N_{NS5}^3 N_{D5}^{3/2}$ \\ & & & \\
\hline
\end{tabular}
\end{center}
\end{table}
\begin{table}
\caption{Results for $AdS_4\times N(1,1)$ and its supergravity duals.}
\label{N11table}
\begin{center}
\begin{tabular}{| c || c | c | c |}
\hline & & & \\
Background & \begin{tabular}{c}Page Charge\\ [4pt] Cycle\end{tabular} & `Raw' CC & CC \\ & & &  \\
\hline & &  & \\
$AdS_4\times N(1,1)$ & \begin{tabular}{c}$Q_{M2}=\left(\frac{\text{R}}{l_p}\right)^6\frac{3}{16\pi^2}=N_{M2}
$\\[4pt] \\  $\Sigma_7=[\mu,\alpha,\beta,\theta_1,\phi_1,\theta_2,\phi_2]$\end{tabular} & $\frac{\pi^4}{2\sqrt{2}}\left(\frac{\text{R}}{l_p}\right)^9$ &$\sqrt{\frac{2}{3}}\frac{16\pi^7}{3}N_{M2}^{3/2}$  \\ & & & \\
\hline & & & \\
$\Downarrow_{\beta}(AdS_4\times N(1,1))$ & \begin{tabular}{c}$Q_{D2}=\frac{\text{R}^9}{l_p\alpha'^{5/2}}\frac{3}{64\pi^2}=N_{D2}$\\[4pt] \\ $\Sigma_6=[\mu,\alpha,\theta_1,\phi_1,\theta_2,\phi_2]$\end{tabular} & $\frac{\pi^3}{16\sqrt{2}}\frac{\text{R}^9}{l_p\alpha'^4}$ & $\sqrt{\frac{2}{3}}\frac{16\pi^6}{3}\frac{\sqrt{l_p}}{\alpha'^{1/4}}N_{D2}^{3/2}$ \\ & & &  \\
\hline & & & \\
NATD($\Downarrow_{\beta}(AdS_4\times N(1,1))$) & \begin{tabular}{c}$Q_{D5}=\frac{\text{R}^6}{l_p\alpha'^{5/2}}\frac{3}{128\pi}=N_{D5}
$\\[4pt] \\ $\Sigma_3=[\mu,\theta_2,\phi_2]$\end{tabular} & $\frac{m_2^3\pi^5}{192\sqrt{2}}\frac{R^9}{l_p\alpha'^4}N_{NS5}^3$ & $\frac{16\pi^{13/2}m_2^3}{9\sqrt{3}}\frac{\sqrt{l_p}}{\alpha'^{1/4}}N_{NS5}^3 N_{D5}^{3/2}$ \\ & & & \\
\hline
\end{tabular}
\end{center}
\end{table}
\end{landscape}


\section{Discussion and Conclusions}\label{Sec:conclusions}

In this manuscript we have used Non-Abelian T-duality to produce new supergravity backgrounds. We have focused on solutions that preserve an AdS${}_4$ factor. These solutions are particularly relevant in the context of the AdS/CFT correspondence since they describe strongly coupled conformal field theories in three dimensions.
One of the prominent members of the class we consider as a seed case is related to the ABJM field theory.  Another interesting example was provided by the $N(1,1)$ spaces, where the NATD apparently destroyed all of the isometries, other than the isometries of $AdS$ and a residual U(1) isometry.  We expect that these new backgrounds will ultimately enrich the number of entries in the AdS/CFT dictionary.

We have only briefly mentioned the supersymmetry in the case of NATD($\Downarrow_{\tau}(AdS_4\times S^7/\mathbb{Z}_k)$), which we have argued preserves $\mathcal{N}=2$ supersymmetry.  In the other cases, we propose that the reductions along the Hopf fibre coordinate, $\tau$, preserve supersymmetry.   In light of the recent work \cite{Kelekci:2014ima} on supersymmetry and SU(2) NATD we argue that the NATD backgrounds also preserved some fraction of supersymmetry. The main result in \cite{Kelekci:2014ima} shows that, at least for backgrounds with Bianchi IX symmetry, supersymmetry is preserved if the Killing spinors of the original background do not depend on the $SU(2)$ isometry directions. A more covariant way to re-state the independence of the Killing spinor on certain coordinates is the Kosmann derivative; this is akin to the Lie derivative  being the covariant way of stating that the metric has some invariance. For the reduction along $U(1)$'s other than the Hopf fiber we do not have a general geometric argument and relied on some explicitly known cases. It would be interesting to systematically and explicitly study the supersymmetry of the resulting backgrounds.

One interesting open problem would be to track, on the field theory side, the effects of various NATD's. In particular, the analysis we have presented of the holographic central charges seems to point to marked differences when one considers reducing along the Hopf fiber or along some other, susy preserving, $U(1)$ direction. The implications for the dual quiver field theories and its potential cascading phases are only hinted at in the gravity side and it would be quite interesting to explore those phases.
\section*{Acknowledgments}
We are particularly grateful to Carlos N\'u\~nez for many discussions and clarifications and, particularly, for sharing \cite{Macpherson:2015tka} before publication with us. We are also thankful to Niall Macpherson, Fernando Quevedo and Dimitrios Tsimpis for comments and encouragement. LAPZ acknowledges Wolfgang M\"uck and the hospitality of INFN Sezione di Napoli at Universit\'a degli Studi di Napoli Federico II  during the conclusion of this work. This research was supported in part by the National Science Foundation under Grant Nos. 1067889 (University of Iowa).  C.A.W. was partially supported by the National Research Foundation of South Africa NITHP. This research is also supported by funds provided by the Abdus Salam ICTP.

\appendix

\section{Review of NATD}
\label{NATDReview}
In this Appendix we will briefly review the procedures for applying Non-abelian T-dualities for backgrounds admitting SU(2) and SU(2)/U(1) isometries.  The procedures are described in more depth in \cite{Itsios:2013wd} for SU(2) and \cite{Lozano:2011kb} for SU(2)/U(1), as well as other coset isometries.
\subsection{SU(2) isometries}
\label{SU(2) isometries}
We follow \cite{Itsios:2013wd} in the generalized 3-step B{\"u}scher procedure and consider backgrounds with an SU(2) isometry such that the metric can be written in the form of
\begin{equation}
\label{generic metric}
ds^2=G_{\mu\nu}(x)dx^{\mu}dx^{\nu}+2G_{\mu i}(x)d
x^{\mu}L^i+g_{ij}(x)L^iL^j,
\end{equation}
where $\mu,\nu = 1,...7$  and $i,j=1,2,3$.  The $L^i$'s are the $su(2)$ Maurer-Cartan forms ($L_{\pm}^i=-i\text{Tr}(t^ig^{-1}\partial_{\pm}g)$, with $g$ an element of $SU(2)$).  All of the Type IIA backgrounds considered in this paper have $B_2=0$, so for simplicity we omit its contribution from the general procedure. The Lagrangian density for the NS sector fields (omitting the dilaton contribution) is given by,
\begin{equation}
\label{L0short}
\mathcal{L}_0=Q_{AB}\partial_{+} X^A\partial_{-}X^B,
\end{equation}
where $A,B=1,...,10$ and
\begin{equation}
Q_{AB}=
\left(
\begin{array}{c|c}
\\ \ Q_{\mu\nu}\quad & Q_{\mu i} \\ \\  \hline \
 Q_{i\mu} \quad  & E_{ij} \\
\end{array}
\right),
\quad  \text{and}\quad  \partial_{\pm}X^A=\left(\partial_{\pm}X^{\mu},\ L_{\pm}^i\right),
\end{equation}
with \begin{equation}
Q_{\mu\nu}=G_{\mu\nu},\quad Q_{\mu i}=G_{\mu i},\quad Q_{i\mu}=G_{i \mu},\quad E_{ij}=g_{ij}.
\end{equation}
We then gauge the SU(2) isometry by changing derivatives to covariant derivatives and introduce gauge fields, $A_{\pm}$ according to $\partial_{\pm}g\to D_{\pm}g=\partial_{\pm}g-A_{\pm}g$. The next step is to add a Lagrange multiplier term to Eq. (\ref{L0short}) to ensure that the gauge fields have vanishing field strength.  The Lagrange multiplier term is given by,
\begin{equation}
-i\text{Tr}(\alpha'vF_{\pm}),\quad F_{\pm}=\partial_+A_--\partial_-A_+-[A_+,A_-].
\end{equation}
Since the dimension of $SU(2)$ is three, we have introduced three new dynamical variables in the form of the Lagrange multipliers, $v_i$, so we must eliminate three of the variables by making a gauge fixing choice.  A natural choice is $g=\mathbb{I}$, so that all of the Euler angles in the SU(2) are zero and all three of the Lagrange multipliers become dual coordinates.
The last step is to integrate out the gauge fields to obtain the dual Lagrangian density,
\begin{equation}
\hat{\mathcal{L}}=\hat{Q}_{AB}\partial_{+} \hat{X}^A\partial_{-}\hat{X}^B \label{dualL},
\end{equation}
where we can read off the dual components of $\hat{Q}_{AB}$ from,
\begin{equation}
\hat{Q}_{AB}=
\left(
\begin{array}{c|c}
\\ \ Q_{\mu\nu}-Q_{\mu i}M_{ij}^{-1}Q_{j\nu}\quad & Q_{\mu j}M_{ji}^{-1} \\ \\  \hline \
 -M_{ij}^{-1}Q_{j\mu} \quad  & M_{ij}^{-1} \\
\end{array}
\right),
\quad  \text{and}\quad  \partial_{\pm}\hat{X}^A=\left(\partial_{\pm}X^{\mu},\ \partial_{\pm}v^i\right).
\end{equation}
We have additionally defined:
\begin{equation}
\label{Mdef}
M_{ij}=E_{ij}+f_{ij},\quad \text{with}\quad f_{ij}=m_2 \alpha'\e_{ij}^{\ \ k}v_k.
\end{equation}
We can identify the dual metric and the generated $\hat{B}_2$ field as the symmetric and antisymmetric components of $\hat{Q_{AB}}$, respectively. The transformation of the dilaton is given by
\begin{equation}
\label{dualdilaton}
\hat{\Phi}=\Phi-\frac{1}{2}\text{ln}(\frac{\text{det}M}{\alpha'^3}).
\end{equation}
In order to transform the RR Fluxes, one must construct a bispinor out of the RR forms and their Hodge duals, (in Type IIA):
\be
P=\frac{e^{\Phi}}{2}\sum^5_{n=0}\slashed{F}_{2n},
\ee
where $\slashed{F}_p=\frac{1}{p!}\Gamma_{\mu_1...\mu_p}F_p^{\ \mu_1...\mu_p}$. The dual fluxes simply arise from inverting $\Omega$:
\be
\hat{P}=P\cdot \Omega^{-1},
\ee
where
\be
\Omega=(A_0\Gamma^1\Gamma^2\Gamma^3+A_a\Gamma^a)\Gamma_{11}/\sqrt{\alpha'^3},\ee and
\be
A_0=\frac{1}{\sqrt{1+\zeta^2}},\quad \quad A^a=\frac{\zeta^a}{\sqrt{1+\zeta^2}}.
\ee
We also need to define $\zeta^a=\kappa^a_{\ i}z^i$ with $\kappa^a_{\ i}\kappa^a_{\ j}=g_{ij}$ and  $z^i=\frac{1}{\det \kappa}(b_i+v_i)$.

\subsection{Coset spaces: $SU(2)/U(1)$ isometry}
\label{Coset Review}
Here we present a quick review of performing the dualization procedure on a background with an SU(2)/U(1) isometry.  A more complete description of generating the Non-Abelian T-dual of a background with coset space isometries was first presented in \cite{Lozano:2011kb}.
We start with a metric of the form of Eq. (\ref{generic metric}), where $\mu, \nu=1,...8$ and $i,j=1,2$, and the $L^i$'s are replaced with $L_1=d\theta, \ L_2=\sin\theta d\phi$.  The procedure is essentially the same as in Appendix \ref{SU(2) isometries} above, but we modify,
\be
E_{ij}=\text{diag}(g_{ij},\lambda),
\ee
where $g_{ij}$ is 2-dimensional.  Due to this, we must additionally gauge fix one of the three Lagrange multipliers.  In all of the cases considered, we choose $(v_1=0, v_2=m_2\alpha' \rho\cos\chi, v_3=m_2\alpha' \rho\sin\chi)$.  After inverting $M_{ij}$, we take $\lambda\to 0$ and construct the dual frames from
\be
\hat{e}=\kappa M^{-T} dv.
\ee

\section{Review of Supergravity EOM}
\label{ReviewEOM}
In this Appendix we will briefly review the 11 and 10D supergravity equations of motion, presented in the convenient form of \cite{Itsios:2013wd,Barranco:2013fza}.  The 11D supergravity action is given by \cite{Cremmer:1978km},
\be
S_{11D}=\frac{1}{\kappa_{11}^2}\int_{M_{11}}\sqrt{g}\big[R-\frac{1}{2}\big(\frac{F_4^2}{4!}+F_4\wedge F_4\wedge C_3\big)\big]
\ee
The 11D Einstein's equations and Bianchi identities can be expressed as,
\bea
&R_{AB}=\frac{1}{12}(F_{4}^{\ 2})_{AB}-\frac{1}{6}g_{AB}\frac{1}{4!}(F_4^{\ 2})_{AB},\\&
d\star F_4=\frac{1}{2}F_4\wedge F_4,\quad dF_4=0.
\eea
The action of massive Type IIA supergravity in string frame without sources is given by,
\bea
S_{IIA}&=&\frac{1}{\kappa^2}\int_{M_{10}}\sqrt{g}\Big[e^{-2\Phi}\bigg(R+4(\partial\Phi)^2-\frac{H^2}{12}\bigg)-\frac{1}{2}\bigg(F_0^2+\frac{F_2^2}{2}+\frac{F_4^2}{4!}\bigg)\Big]\nn \\&~&-\frac{1}{2}\bigg(dC_3\wedge dC_3\wedge B_2 +\frac{F_0}{3}dC_3\wedge B_2^3+\frac{F_0^2}{20}B_2^5\bigg).
\label{SIIA}
\eea
In this case, the Bianchi identities are given by,
\bea
dF_0=0,\ \ dF_2=F_0H_3,\ \ dF_4=H_3\wedge F_2,
\eea
which can be deduced from the definitions of the fluxes,
\be
F_0=m,\ \ F_2=dC_1+F_0 B_2,\ \ F_4=dC_3-H_3\wedge C_1+\frac{F_0}{2}B_2\wedge B_2.
\ee
The equations of motion that follow from varying (\ref{SIIA}) with respect to the metric are,
\bea
R_{\mu\nu}+2D_{\mu}D_{\nu}\Phi=\frac{1}{4}H_{\mu\nu}^2+e^{2\Phi}\Big[\frac{1}{2}(F_2^2)_{\mu\nu}+\frac{1}{12}(F_4^2)_{\mu\nu}-\frac{1}{4}g_{\mu\nu}\bigg(F_0^2+\frac{1}{2}F_2^2+\frac{1}{4!}F_4^2\bigg)\Big],\nn \\
\eea
and the dilaton equation is,
\be
R+4D^2\Phi-4(\partial\Phi)^2-\frac{1}{12}H^2=0.
\ee
The action of Type IIB supergravity in string frame without sources is given by,
\bea
S_{IIB}&=&\frac{1}{\kappa^2}\int_{M_{10}}\sqrt{g}\Big[e^{-2\Phi}\bigg(R+4(\partial\Phi)^2-\frac{H^2}{12}\bigg)-\frac{1}{2}\bigg(F_1^2+\frac{F_3^2}{3!}+\frac{1}{2}\frac{F_5^2}{5!}\bigg)\Big]\nn \\&~&-\frac{1}{2}C_4\wedge H\wedge dC_2
\label{SIIB}
\eea
The Bianchi identities are an additional constraint on the fluxes given by,
\be
dH=0,\ \ dF_1=0,\ \ dF_3=H_3\wedge F_1,\ \ dF_5=H\wedge F_3.
\ee
Here, the definition of the fluxes take the form,
\be
H_3=dB_2,\ \ F_1=dC_0,\ \ F_3=dC_2-C_0 H_3,\ \ F_5=dC_4-H\wedge C_2.
\ee
The equations of motion that follow from varying (\ref{SIIB}) with respect to the metric are,
\bea
&R_{\mu\nu}+2D_{\mu}D_{\nu}\Phi=\frac{1}{4}H_{\mu\nu}^2+e^{2\Phi}\Big[\frac{1}{2}(F_1^{2})_{\mu\nu}+\frac{1}{4}(F_3^{2})_{\mu\nu}+\frac{1}{96}(F_5^{2})_{\mu\nu}-\frac{1}{4}g_{\mu\nu}\bigg(F_1^{2}+\frac{1}{3!}F_3^{2}\bigg)\Big]\nn \\&
\eea
and the dilaton equation is again given by,
\be
R+4D^2\Phi-4(\partial\Phi)^2-\frac{1}{12}H^2=0.
\ee


\bibliographystyle{utphys}
\bibliography{Ypqbib}

\end{document}